\documentclass[12pt]{iopart}

\pdfoutput=1 %required on archive
\usepackage{graphics}      % standard graphics specifications
\usepackage{graphicx}      % alternative graphics specifications
\usepackage{slashed}
\usepackage{subfigure}
\usepackage{float}
\usepackage{longtable}     % helps with long table options
\usepackage{url}           % for on-line citations
\usepackage{bm}            % special 'bold-math' package 

\usepackage{iopams}
\newcommand{\I}{\mathrm{I} }
\newcommand{\II}{\mathrm{II}}
\newcommand{\III}{ \mathrm{III}}

\begin{document}

\title[Anomalous Currents on Closed Surfaces]{Anomalous Currents on Closed Surfaces: \\ Extended Proximity, Partial Quantization, and Qubits}

\author{Alexander Selem}

\address{Department of Physics, University of California, Berkeley, CA 94720, USA}
\address{Berkeley Quantum Information and Computation Center, University
of California, Berkeley, CA 94720, USA}
\ead{aselem@berkeley.edu}

\begin{abstract}
Motivated by the surface of topological insulators, the Dirac anomaly's discontinuous dependence on sign of the mass, $m/|m|$, is investigated on closed topologies when mass terms are weak or only partially cover the surface.  It is found that, unlike the massive Dirac
theory on an infinite plane, there is a smoothly decreasing current when the mass region is not
infinite; also, a massive finite region fails to exhibit a Hall current edge—exerting an extended
proximity effect, which can, however, be uniformly small—and oppositely orientated Hall
phases are fully quantized while accompanied by diffuse chiral modes.
Examples are computed using Dirac energy eigenstates on a flat torus (genus one topology) and closed cap cylinder (genus zero topology) for various mass-term geometries. 
Finally, from the resulting the properties of the surface spectra, a potential application for a flux-charge qubit is presented. 
\end{abstract}

\date{\today}

\maketitle

\section{Introduction}
The 2+1 dimensional massive Dirac theory on an infinite plane exhibits an anomalous contribution to the Hall conductance ($\hbar, c,=1$, $e=|e|$) $\sigma_{xy} = \frac{m}{|m|} \frac{e^2}{4\pi}$~\cite{Jackiw:1984ji, Niemi:1983rq, Redlich} representing a \emph{half} charge pumped contribution per unit of threaded flux in a Laughlin type setup~\cite{PhysRevB.23.5632, PhysRevB.25.2185}. Unfortunately, this half quantization is not directly observable, since a ﬂat two-dimensional lattice theory such as that for graphene, must have an even number of Dirac modes~\cite{Nielsen1983389, PhysRevB.73.125411, PhysRevLett.95.146801}. Having effectively two or more pairs of fermion flavors, single fermion results are effectively doubled. 

In contrast, the surface of 3-dimensional topological insulators contain an odd number of massless Dirac fermion modes on closed topologies~\cite{PhysRevLett.98.106803, PhysRevB.76.045302, Bernevig15122006, PhysRevLett.96.106802, Konig02112007, hsieh-2008-452, xia-2009-5, Zhang_Liu_Qi_Dai_Fang_Zhang_2009, PhysRevB.78.195424, PhysRevLett.103.196804}.  Effective Dirac mass terms could be given to the surface theory at selective regions, by coupling to a ferromagnetic time-reversal breaking material~\cite{Chen06082010}. An application of the naive infinite plane results to this system suggests several questions. 
If the massive region only covers a small area, will charge accumulate on its edge? If not, can a sign change of a weak local mass induce a sign flipping transition of the quantized current globally?  Finally, how do  magnetic fluxes induce fractional charge pumping in specific cases? In the present work these issues are investigated and clarified.

Rosenberg \emph{et al}~\cite{Rosenberg} gave a solution to the last question by concluding that fractional charge accumulation did not occur when a flux is inserted through a topological insulator. The surface theory was said to breakdown at $\pi$ flux, as gapless bands induced in the bulk were responsible for discharging the ends of the flux tube.  This was called the wormhole effect. Yet, it is not possible to explain a half-charge quantized transport on the basis of bands. In fact, the two dimensional plane result implies the existence of the current in a fully gapped theory. The results to be presented will clarify these points and it is conjectured that fractional charge could indeed be observed.

In the present work it is emphasized that the surface theory is sufficient to resolve all previous questions.  Three results will be shown through solved examples and motivated from the effective action of topological insulators. These are: 1) that mass terms do not induce perfect quantization in weak limits for closed finite surfaces (partial quantization), 2) that mass edges do not result in fractional Hall conductance edges (``extended proximity" effect) and 3) when the transition region separates oppositely orientated masses those phases are fully quantized but overlapped by a sufficiently diffuse chiral band. The existence of chiral bands is not a new result~\cite{PhysRevLett.103.196804, PhysRevB.84.085312, 0953-8984-24-1-015004}, but rather the full quantization of the separated phases. It is also suggested that the wormhole effect~\cite{Rosenberg} only occurs in the localized flux limit and in that case can be thought of as an example of the extended proximity effect on a genus one surface.

Section~\ref{sec:conj} will first offer motivation for these results from the effective surface action. Secs.~\ref{sec:torus} and \ref{sec:closed} will then present specific examples and constructions showing how the effects are manifested by the surface Dirac theory.
The solutions to the Dirac equation has been investigated in other contexts and geometries involving topological insulators and graphene~\cite{PhysRevLett.103.196804, Gonzalez1993771, PhysRevB.83.075424, PhysRevB.76.165409}. 
The anomalous Hall current is understood from the wavefunction spectral asymmetry (see~\ref{sec:anom}). Finally, gaining intuition from the different mass geometries computed, a potential configuration of masses with inserted flux is proposed as an architecture for a qubit in section~\ref{sec:qubit}.

%%%%%%%%%%%%%%%%%%%%%%%%%%%%%%%%%%%%%%%%%%%%%%%%%%%%%%%%%%%%%%%%%%%%%%%%%%%%
%%%%%%%%%%%%%%%%%%%%%%%%%%%%%%%%%%%%%%%%%%%%%%%%%%%%%%%%%%%%%%%%%%%%%%%%%%%%
\section{Effective Action} %ummary of results and conjectures
\label{sec:conj}

The conclusions reached in this work can be motivated from the effective field theory for topological insulators. Starting from the topological BF theory it was shown in reference~\cite{Moore} that the entire electromagnetic response comes from the surface.
After integrating out the effective fields, the Hall action is obtained:
\begin{equation}
S_{surf}= -\frac{e^2}{8\pi^2}\int \epsilon^{\mu \nu \rho} \theta A_\mu\partial_\nu A_\rho.
\label{eq:hall}
\end{equation}
Here, and throughout, Greek indices take values over the 2+1 space-time describing the surface. $\theta$ is a parameter or background field giving a hall conductance $\sigma_{xy}=  \theta e^2/{4\pi^2}$. $\theta$ is required to be $\pm \pi$ to respect time-reversal invariance, and comparing with the Dirac theory on a plane allows one to identify $\theta =$Sign$(m) \pi$~\cite{Jackiw:1984ji, Niemi:1983rq, Redlich}. 

\emph{Partial Quantization.} First it is noted that on the surface of the insulator there is no reason to expect $\theta$ to be quantized if time-reversal invariance is being broken by mass-terms.  Importantly, for the Abelian theory on a closed surface (as in equation~\ref{eq:hall}), gauge invariance is consistent with $\theta$ unquantized. A fixed value of $\theta$ other than $\pm \pi$ would lead to a an unquantized Hall current. 

\emph{Extended Proximity Effect.} For a more complicated arrangement of mass terms it might be allowed $\theta \rightarrow \theta(x)$, as an effective parameter not necessarily equivalent to the local sign of the mass.  An abrupt change in the value of $\theta$ from non-zero to zero can be thought of as a Hall edge. However, gauge invariance in general requires that $\theta$ remain constant:
\begin{equation}
S_{surf} \rightarrow -\int \epsilon^{\mu\nu\rho}\theta\partial_\mu \Lambda \partial_\nu A_\rho = \int \epsilon^{\mu\nu\rho}\Lambda \partial_\mu \theta  \partial_\nu A_\rho \neq 0,
\label{eq:BFG}
\end{equation}
($0$ mod $2\pi$ implied for the last term). Therefore $\theta \neq \theta(x)$ which is presumably responsible for the extended proximity effect and the absence of a Hall edge.

\emph{Quantization of opposite phases.} The exception to equation~\ref{eq:BFG} is if a region exists of zero magnetic field (which is the case for closed surfaces that do not surround monopoles). In these cases $\theta$ can transition (over a region of zero field) from one fully quantized (half-charge) Hall phase to the oppositely quantized phase accompanied by a chiral band.

To see this, working in the Coulomb gauge ($A_0 =0$), zero field strength requires $A_i=\partial_i \Gamma$ ($i  \in 1, 2$).  Inserting this into the action where $\theta$ varies, naively gives:
\begin{equation}
S_{\theta\neq \theta_o}= \frac{e^2}{4\pi^2}\int \epsilon^{i j} \partial_i \theta ( \partial_0 \Gamma \partial_j \Gamma) .
\end{equation}
Normally the electromagnetic field is physically determined by a single choice of $\Gamma$. Then there are no unique modes and no way to form propagating packets. However on the compact surface there exist different disjoint $\Gamma$ distinguished by winding number. These different winding numbers can act as different modes forming a compact (discrete) chiral band.

If $\theta$ varies in the $j$ direction, the $i$-direction is topologically equivalent to a circle separating the two phases, and with periodic boundary conditions in time, the action can be integrated to give (setting $e=1$):
\begin{equation}
S_{\theta\neq \theta_o}= \Delta \theta k l 
\end{equation}
where $k$, $l$, are integers. This remains gauge invariant for $\Delta \theta = 2 \pi n$ ($n \in Z$). Therefore $\theta$ must separate two fully quantized oppositely orientated phases. The complete picture is that the chiral band can absorb or release a whole charge, half of which comes from each separate phase. The modes may be diffuse over the entire region where the field strength vanishes, however.

In the rest of this work, these suggested results will be shown in various solvable examples.

%%%%%%%%%%%%%%%%%%%%%%%%%%%%%%%%%%%%%%%%%%%%%%%%%%%%%%%%%%%%%%%%%%%%%%%%%%%%
%%%%%%%%%%%%%%%%%%%%%%%%%%%%%%%%%%%%%%%%%%%%%%%%%%%%%%%%%%%%%%%%%%%%%%%%%%%%
\section{Flat Torus}
\label{sec:torus}
The first easily solvable case is a flat torus, which can be considered as a simplification for a crystal wafer with a hole drilled through it. 
Labeling coordinates $z$ and $\phi$ for the two orthogonal directions on the torus, a flat metric can be used. It is important to note that four inequivalent forms exist for the Dirac equation on the torus, corresponding to the inequivalent spin structures. For example writing the torus as a square with edges identified naturally leads one to a Cartesian-like form (with constant Pauli matrices). However, this will not agree from an embedding in 3-dimensional Euclidian space as used in references~\cite{Zhang, 2010arXiv1005.3542Z, Rosenberg}. While both forms are mathematically consistent, they lead to different physical results.  The spectrum, for example, has a finite-size gap present in one case but not in the other. However, none of the conclusions made here will depend on this choice. Therefore I will pick the spin structure that reproduces the half-integer azimuthal dependence found for the unique choice in the case of a \emph{closed} (with caps) cylinder (see section~\ref{sec:closed} and~\ref{sec:sol_torus}). It is also the choice which leads to a derivation of the Dirac equation by embedding.

\subsection{Weak Hall current}
\begin{figure}
\centering
\subfigure{\includegraphics[width=.35\textwidth]{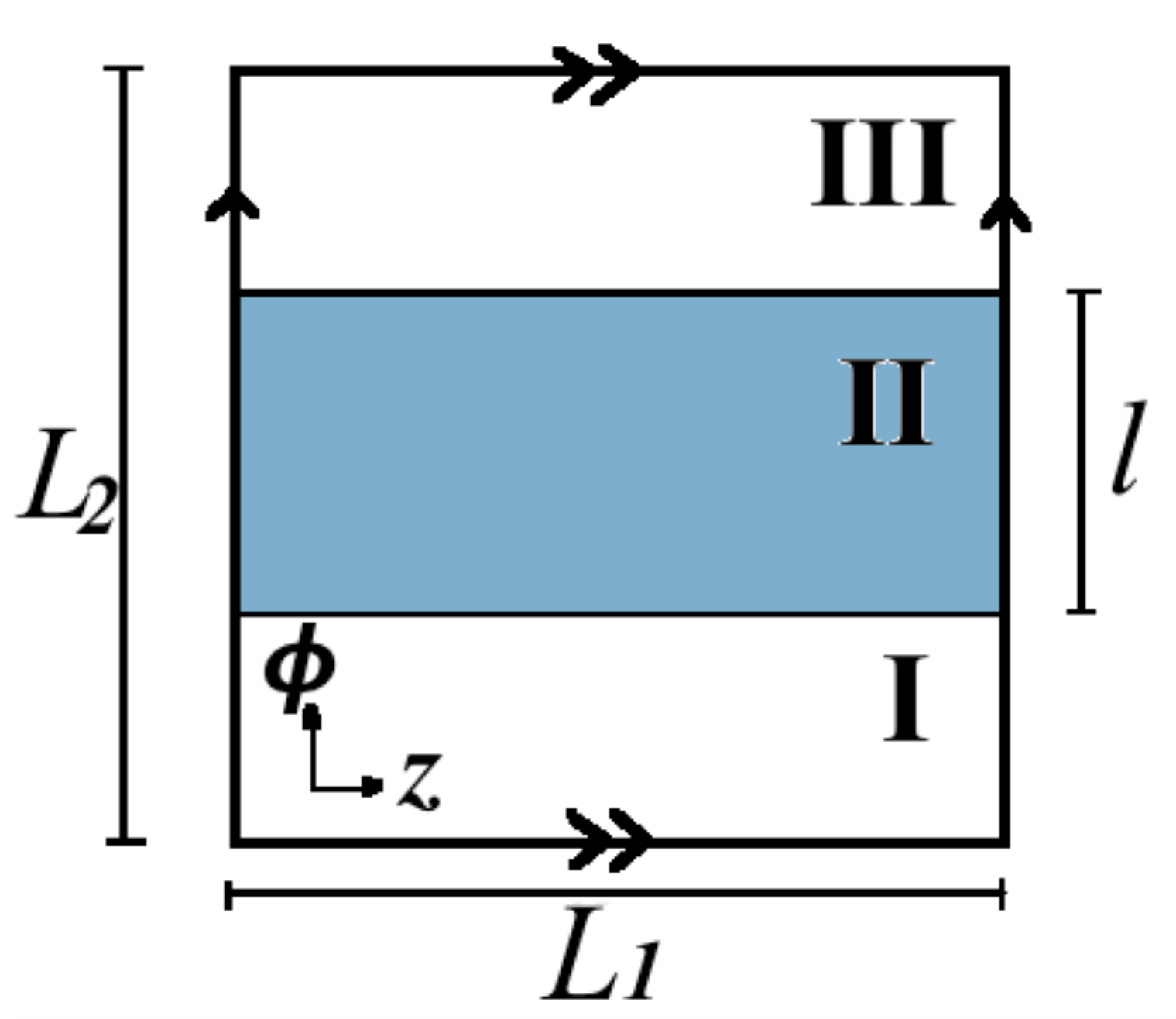}}
\subfigure{\includegraphics[width=.35\textwidth]{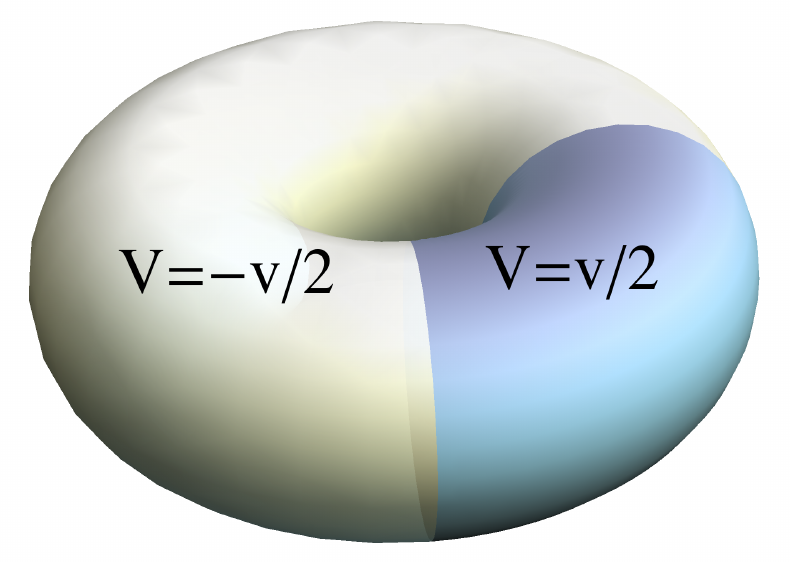}}
\caption{(Color online) Coordinates and dimensions (left) for the smooth torus (right). Note sides are identified in the left figure. The surface is massive (constant) everywhere, but a non-zero scalar step-like potential is introduced with values as shown by $V$ (right). $z$ is in the vertical direction at right. Note the torus is separated into two regions, or three with edges identified, which will be labeled as I, II, III. \label{toruscoords1}}
\end{figure}

\begin{figure}
\centering
\includegraphics[width=.45\textwidth]{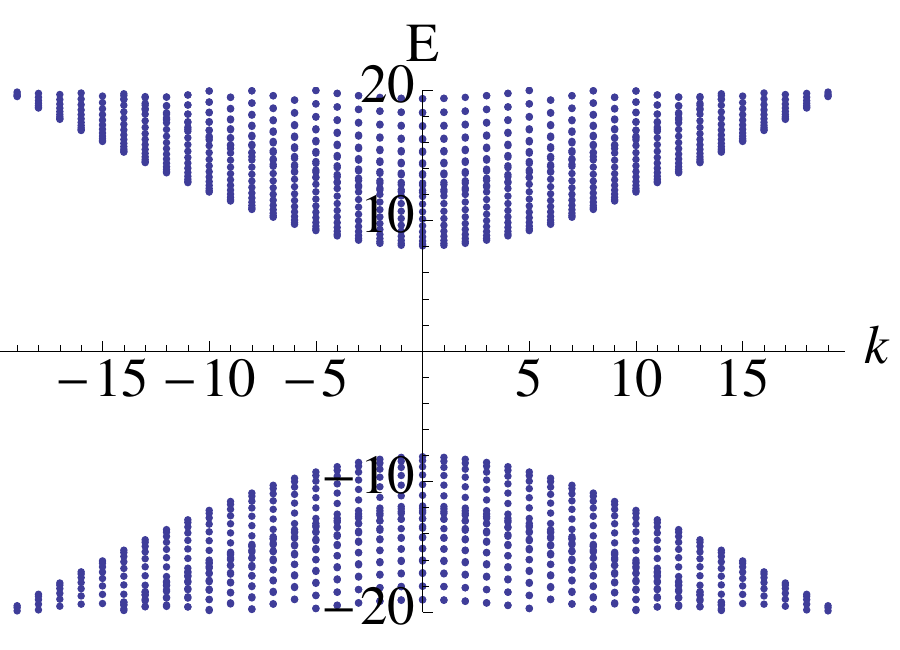}
\caption{(Color online) Spectrum for the configuration of figure~\ref{toruscoords1} with $m=10$, $v=4$, $l=2$, $L_2=5$. Units are in terms of $2\pi/L_1$ as defined in the text. \label{torusEres}}
\end{figure}

To ascertain an anomalous Hall conductance as a function of mass-strength I consider a fully massive Dirac fermion with a step potential in the $\hat{\phi}$ direction $V(\phi)=v(\Theta(\phi+l/2)-\Theta(\phi-l/2))-v/2$, corresponding to localized electric fields at $\pm l/2$, $\vec{E}= [-\delta(\phi+l/2)+\delta(\phi-l/2)]v\hat{\phi}$. This construction is shown in figure~\ref{toruscoords1}. The units are set with $L_1=2\pi$ and other quantities are measured in units of $L_1$, $L_2\rightarrow 2\pi L_2/L_1$, $l\rightarrow 2\pi l/L_1$. Then the Dirac equation is ($\gamma^{0}=\sigma^3, \gamma^{z}=\sigma^3\sigma^1, \gamma^{\phi}=\sigma^3\sigma^2$):
\begin{equation}
\label{torham3}
e^{-i\frac{\sigma^3}{2}\phi}
(-i\sigma^1\partial_{z}-i\sigma^2\partial_{\phi}+\sigma^3m -e V(\phi)
)e^{i\frac{\sigma^3}{2}\phi}\psi
=E\psi.
\end{equation}
with $0 \leq z < 2\pi$ and $-L_{2}/2 \leq \phi < L_{2}/2$ while mass and energy are normalized to units of $2\pi/L_{1}$. The solutions in each region are matched and the quantization condition derived in~\ref{sec:sol_torus}. The corresponding spectrum is unremarkable and shown in figure~\ref{torusEres} as a function of the azimuthal angular momentum. It remains fully gapped for any threaded flux.
\begin{figure}
\centering
\raisebox{-.16cm}{\includegraphics[width=.5\textwidth]{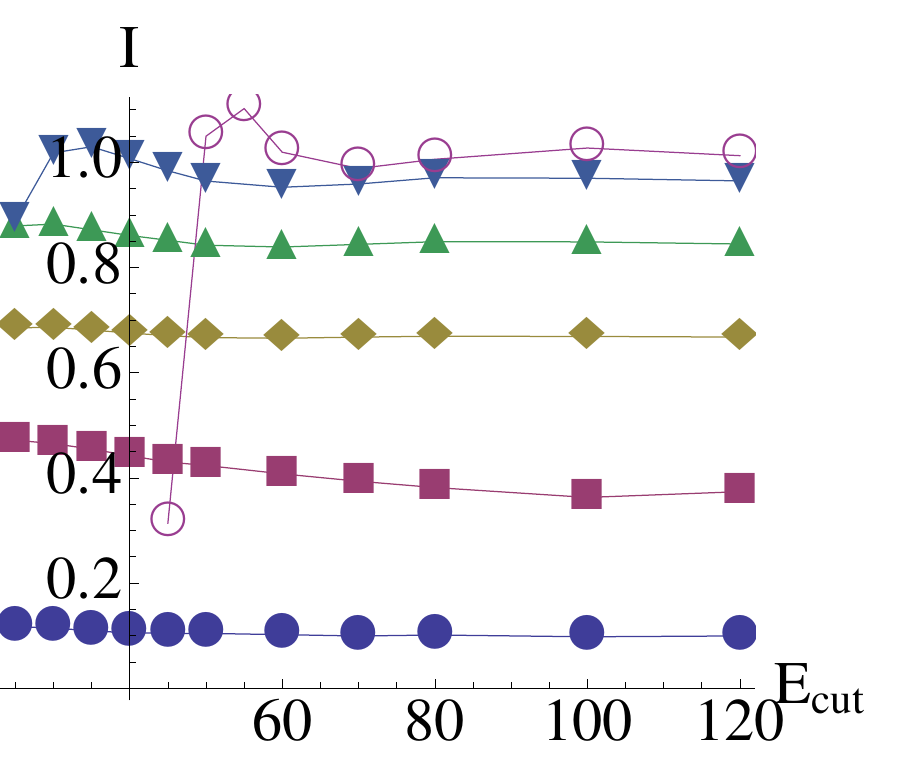}}
\caption{(Color online) Hall current, $I$ (in the $\hat{z}$-direction) in units of $-e/2v$, for the configuration of figure~\ref{toruscoords1} along a strip of width $=L_2/100$ centered around $\phi=-l/2$. From the infinite plane result, $I=1$ expected. The $x$-axis labels the cutoff energy employed (that is, all states below said energy are summed to approximate the delta function peak). Each line represents a different mass: from top to bottom $m=40$, $20$, $10$, $5$, $2.5$, $.5$ (in units of $2\pi/L_1$). All cases take $v=4$, $l=2$, and $L_2=5$. The same values with opposite sign are obtained along the strip $\phi=l/2$. \label{torusrescur}}
\end{figure}

Nevertheless the anomalous Hall current, $j^{z}(\phi)$, is non-zero and can be computed numerically using equation~(\ref{currenti}). According to the infinite plane result, one expects $\frac{j^{z}(\phi)}{-e}= [\delta(\phi+l_{\phi}/2)-\delta(\phi-l_{\phi}/2)]\frac{v}{2} \mathrm{Sign}(m)$. However, the computed result turns out to have a smoother mass dependence as shown in figure~\ref{torusrescur}, illustrating partial quantization. The current smoothly approaches zero as the mass approaches zero with the onset of full quantization only with a sufficiently large mass-distance $m\sim 20$ (in units of $2\pi/L_1$).
If the current is computed along any window other than the delta function peaks, none is found, as expected.  There is no further dependence on the potential strength beyond the expected (linear) scaling. Finally, note that the system exhibits a current despite remaining fully gapped; consistent with the understanding that bands are not responsible for fractional charge transport.

\subsection{Extended proximity effect: first suggestions}
\begin{figure}
\centering
\subfigure{\includegraphics[width=.33\textwidth]{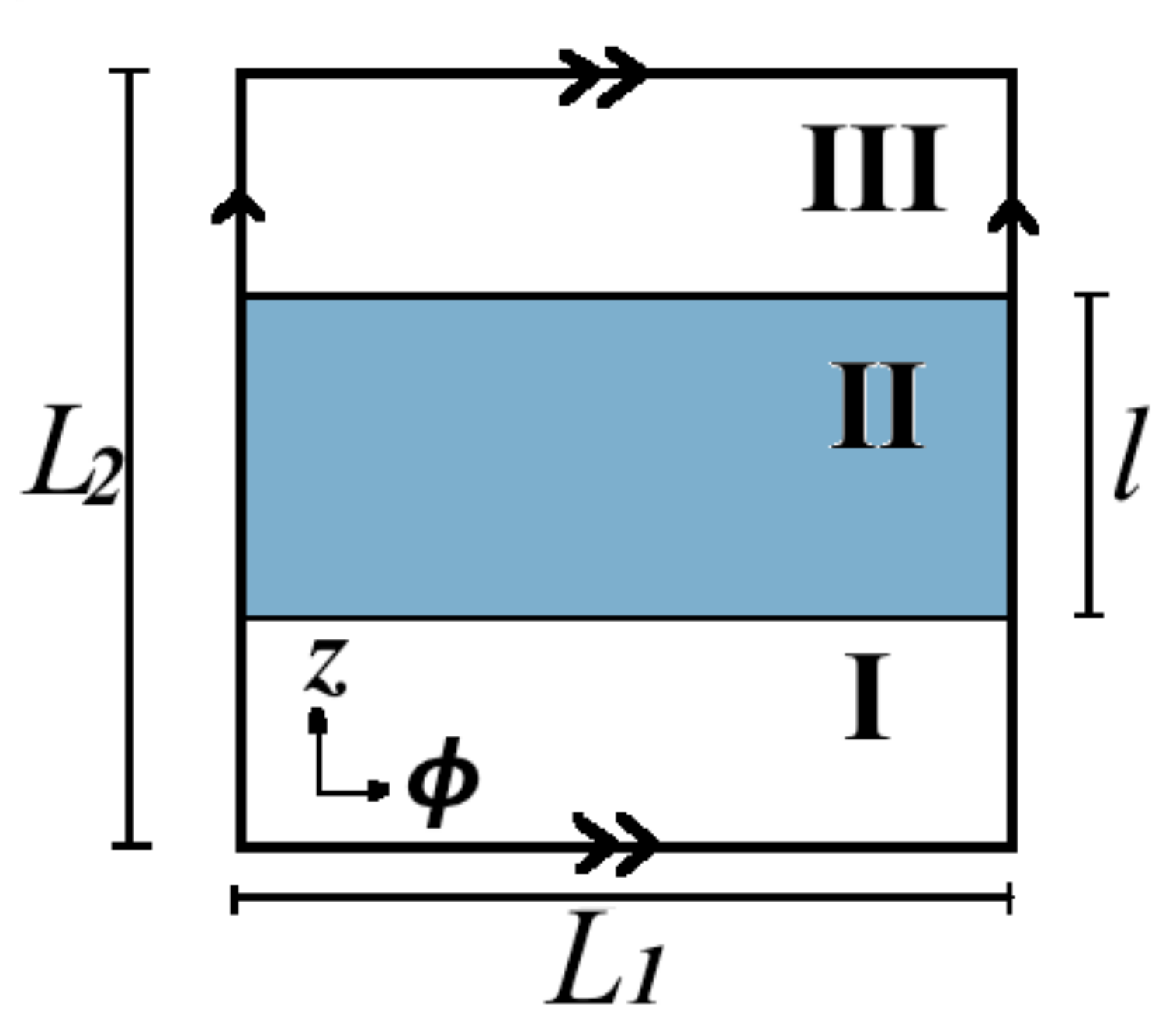}}
\subfigure{\includegraphics[width=.33\textwidth]{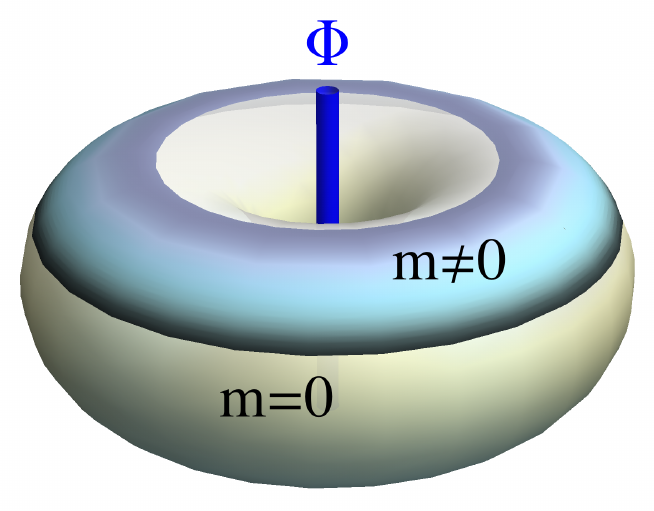}}
\caption{(Color online) Coordinates and dimensions (left) for the smooth torus (right). Note sides are identified in the left figure and $\phi$ and $z$ have been exchanged relative to figure~\ref{toruscoords1}. However, at right, $z$ is still in the vertical direction. The torus is separated into two regions, or three with edges identified, as shown in at left labeled I, II, III. A mass term, $m$, is present only in the shaded area, region II. \label{toruscoords2}}
\end{figure}

A second case that can be easily computed suggests that while the current depends on the mass strength, it does not respond locally to a massive region. To see this, a torus with a strip of mass $m$ is considered as shown in figure~\ref{toruscoords2}. It is checked that no charge accumulates after flux threading, in other words, a strip of mass does not behave like an edge and a Hall phase is globally induced. While only suggestive, much more compelling examples will be considered in the next section.

\begin{figure}
\centering
\includegraphics[width=.33\textwidth]{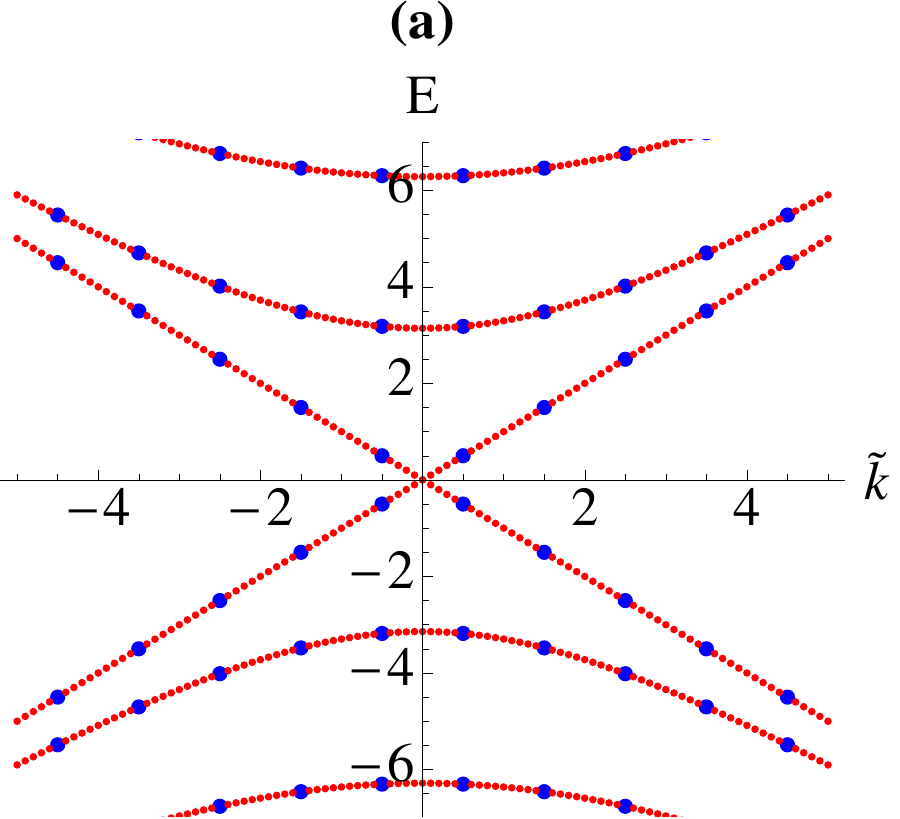} %%fig 6a
\;
\includegraphics[width=.33\textwidth]{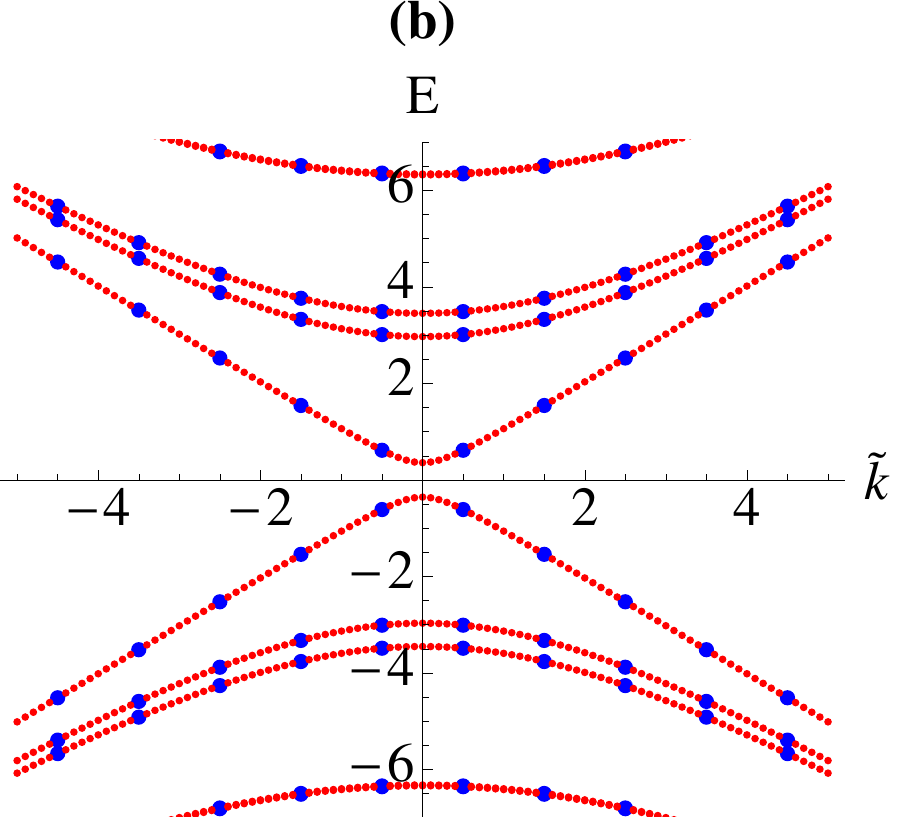}
\caption{(Color online) Spectrum for the configuration of figure~\ref{toruscoords2}. (a) massless case, $m=0$ with $l=.5$ and $L_2=2$. (b) with mass $m=1.5$, and same aspect ratios. Units are in terms of $2\pi/L_{1}$ as described in the text. \label{torusres}}
\end{figure}

The Dirac equation and solutions in this case are given in~\ref{sec:sol_torus} and eigenvalues are shown in figure~\ref{torusres}. The large(blue) points are at the allowed azimuthal quantum numbers $\tilde{k} \in \mathbb{Z}+1/2$ of the wavefunction $\tilde{\psi}$ defined in the appendix as $\tilde{\psi}_{\alpha}=e^{i\frac{\sigma^3}{2}\phi}\psi_{\alpha}$. Because of the half-integer quantization there is a finite-size energy gap, of order $1/L_2$. When a flux $\Phi$ is inserted its effect can be undone through a transformation  which amounts to shifting $\tilde{k}\rightarrow \tilde{k}+\Phi$. The smaller(red) points show the evolution of the eigenvalues under this threaded flux.

For the massless case (figure~\ref{torusres}(a)), the bands traverse the gap.  These bands \emph{are not} responsible for the Hall current and half-integer charge flow around the torus.  The wavefunction for these states satisfy $\psi_{E\leq 0}=\sigma^3 \psi_{E\geq 0} =$constant which does not support a current in the $\hat{z}$ direction or charge accumulation. There is a constant current in the $\hat{\phi}$ direction $=2\pi e^2$, which is the  usual longitudinal current expected of free fermions accelerated under potential $2\pi e$.

For any small strip of mass, a gap forms for all fluxes as shown in figure~\ref{torusres}(b). In this case the flow returns the vacuum to itself, and again no charge accumulation is seen suggestive of a extended proximity effect and lack of Hall edge. This spectral flow is to be contrasted with the results of the next section.

%%%%%%%%%%%%%%%%%%%%%%%%%%%%%%%%%%%%%%%%%%%%%%%%%%%%%%%%%%%%%%%%%%%%%%%%%%%%
%%%%%%%%%%%%%%%%%%%%%%%%%%%%%%%%%%%%%%%%%%%%%%%%%%%%%%%%%%%%%%%%%%%%%%%%%%%%
\section{Surface of a Closed Cylinder}
\label{sec:closed}

The surface of a closed cylinder offers analytic solutions for the more interesting case showing fractional charge accumulation and direct detection of weak quantization and the absence of a Hall edge.

\subsection{Closed Cylinder Cases: weak currents and extended proximity effect}
\label{sec:closedb}

\begin{figure}
\centering
\includegraphics[width=.4\textwidth]{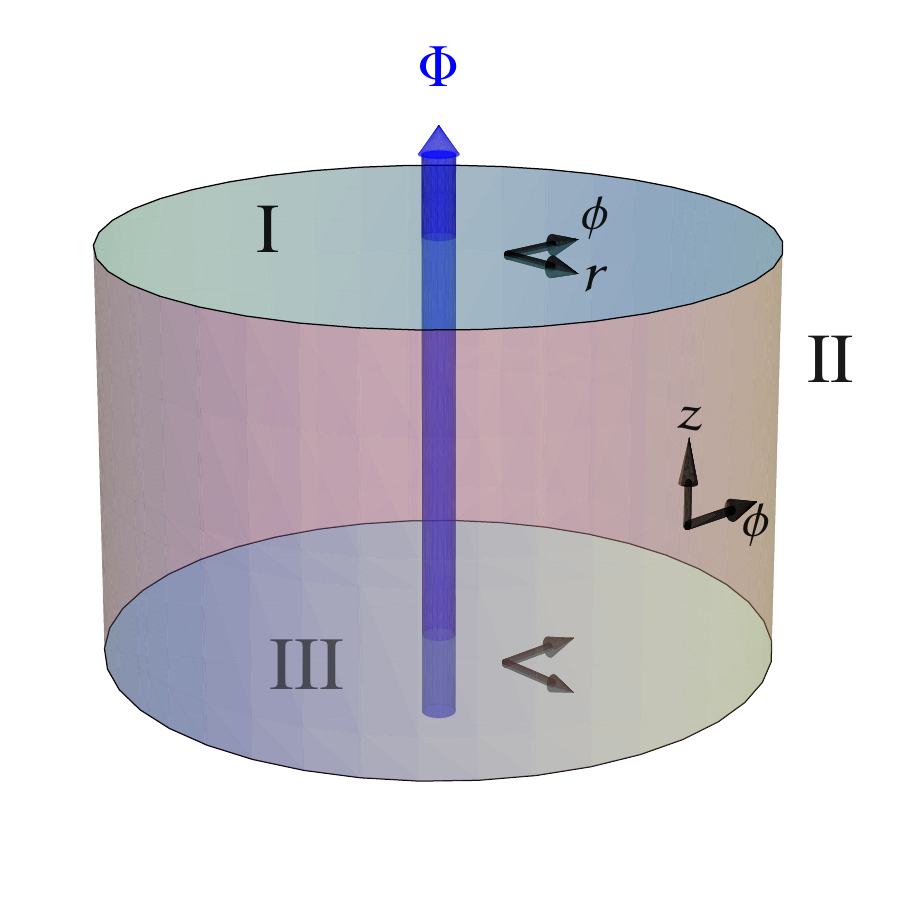}
\caption{(Color online) Closed cylinder of radius $R$, and side (cap-to-cap) length $d$. Three separate sections are labeled I, II, and III each with local coordinates as shown. The caps, I and III, have local polar coordinates with $\hat{r}\times\hat{\phi}$ oriented upward. $\hat{\phi}$ globally has the same orientation. The side has $\hat{z}$ oriented in the upward direction, with the origin set in the center of the side. Different cases are considered where each section may separately have a constant oriented mass $m_{\alpha}$. See text. \label{cyl}}
\end{figure}

The relevant dimensions and coordinates for the cylinder surface are described in figure~\ref{cyl}. 
The Dirac equation in the three different sections and matching conditions are derived in~\ref{sec:eqn_cyl_surf}, they are on the top and bottom caps (with and without masses):
\begin{eqnarray}
\label{cart}
 [(\sigma^{1}\cos\phi & +\sigma^{2}\sin\phi)\partial_{r} +
(\sigma^{2}\cos\phi-\sigma^{1}\sin\phi)\frac{\partial_{\phi}}{r} \\ &+ m_{\I,(\III)}\sigma^3 ]\psi_{\I,(\III)}=E\psi_{\I,(\III)}(r, \phi) 
\end{eqnarray}
and on the side:
\begin{eqnarray}
\label{side}
e^{-i\frac{\sigma^3}{2}\phi}
(-i\sigma^1\partial_{z}-i\sigma^2\partial_{\phi}+\sigma^3m_{\II}
)e^{i\frac{\sigma^3}{2}\phi}\psi_{\II} =E\psi_{\II}(z, \phi) 
,
\end{eqnarray}
or defining in each region $\tilde{\psi}=\exp(i\frac{\sigma^3}{2}\phi)\psi$,
\begin{eqnarray}
\label{diraccyl1}
 \left(-i\sigma^{1}\partial_{r}-i\sigma^{2}\frac{\partial_{\phi}}{r} -\frac{i\sigma^1}{2r}+m_{\I,(\III)}\sigma^3\right)\tilde{\psi}_{\I,(\III)}=E\tilde{\psi}_{\I,(\III)}, 
\\
\left(-i\sigma^1\partial_{z}-i\sigma^2\partial_{\phi}+m_{\II}\sigma^3\right)
\tilde{\psi}_{\II}=E\tilde{\psi}_{\II}. 
\end{eqnarray}
The matching conditions are (see~\ref{sec:eqn_cyl_surf}):
\begin{equation}
\label{diraccyl2}
\tilde{\psi}_{\I}|_{r=R} = \sigma^2\tilde{\psi}_{\II}|_{z=d/2}, \quad 
\tilde{\psi}_{\III}|_{r=R} = \tilde{\psi}_{\II}|_{z=-d/2}.
\end{equation}
I consider six cases ($m>0\neq m(x)$): (a) massless case, $m_{\I}=m_{\II}=m_{\III}=0$ (b) positive mass on top cap alone $m_{\I}=m$, $m_{\II}=m_{\III}=0$, (c) positive mass on the side $m_{\II}=-m$, $m_{\I}=m_{\III}=0$ (d) positive mass on top and bottom $m_{\I}=-m_{\III}=m$, (e) positive mass everywhere $m_{\I}=-m_{\II}=-m_{\III}=m$, and finally oppositely oriented masses on the two caps (f) $m_{\I}=+m_{\III}=m$, $m_{\II}=0$. Note the relative minus sign between the top and the rest of the cylinder in the cases (b)-(e) to describe a \emph{same} sign mass, comes from the $\sigma^2$ transformation in the top-to-side matching; stemming from the opposite orientation of the $r$ and $z$ directions at the top-to-side boundary. 

Unlike the torus, the effect of the flux is not simply to shift the azimuthal quantum number. In particular some boundary condition must be implemented at the origin which physically involves the flux details. Interestingly a simple shift alone of $\tilde{k}$ (azimuthal quantum number of $\tilde{\psi}$) would allow for extra solutions, unconstrained by normalizability at the origin (see reference~\cite{Alford1989140}). The simplest possibility is to take a profile for the flux as a localized delta-function-ring with small radius, $\epsilon$: $B(r)=\Phi\delta(r-\epsilon)/(2\pi\epsilon)$ ~\cite{Alford1989140}, which is expected to be qualitatively similar to a diffuse flux of width $\epsilon$. 

\begin{figure} %star for wide format
\centering
\subfigure{\includegraphics[width=.39\textwidth]{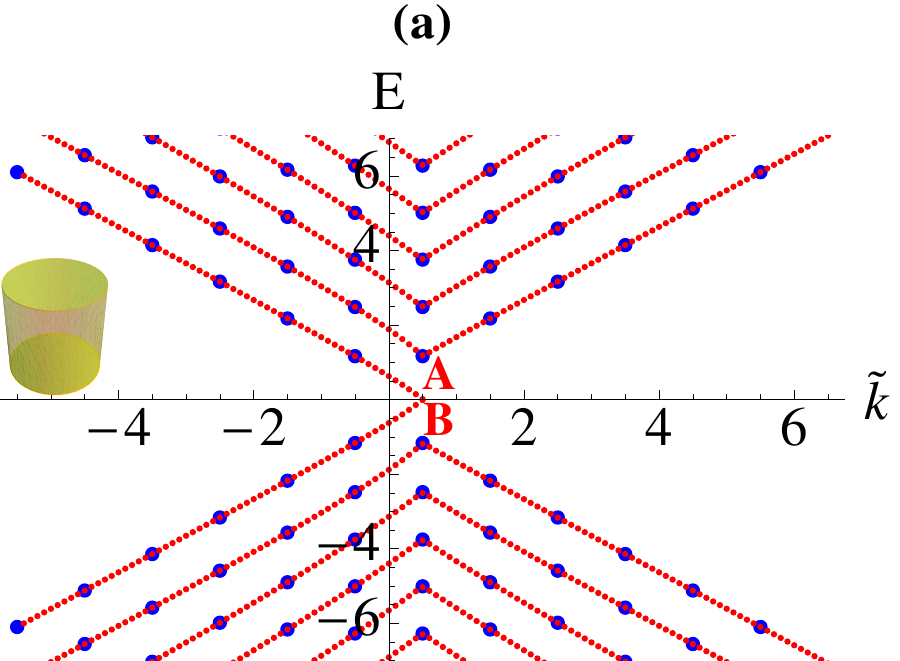}}
\quad
\subfigure{\includegraphics[width=.39\textwidth]{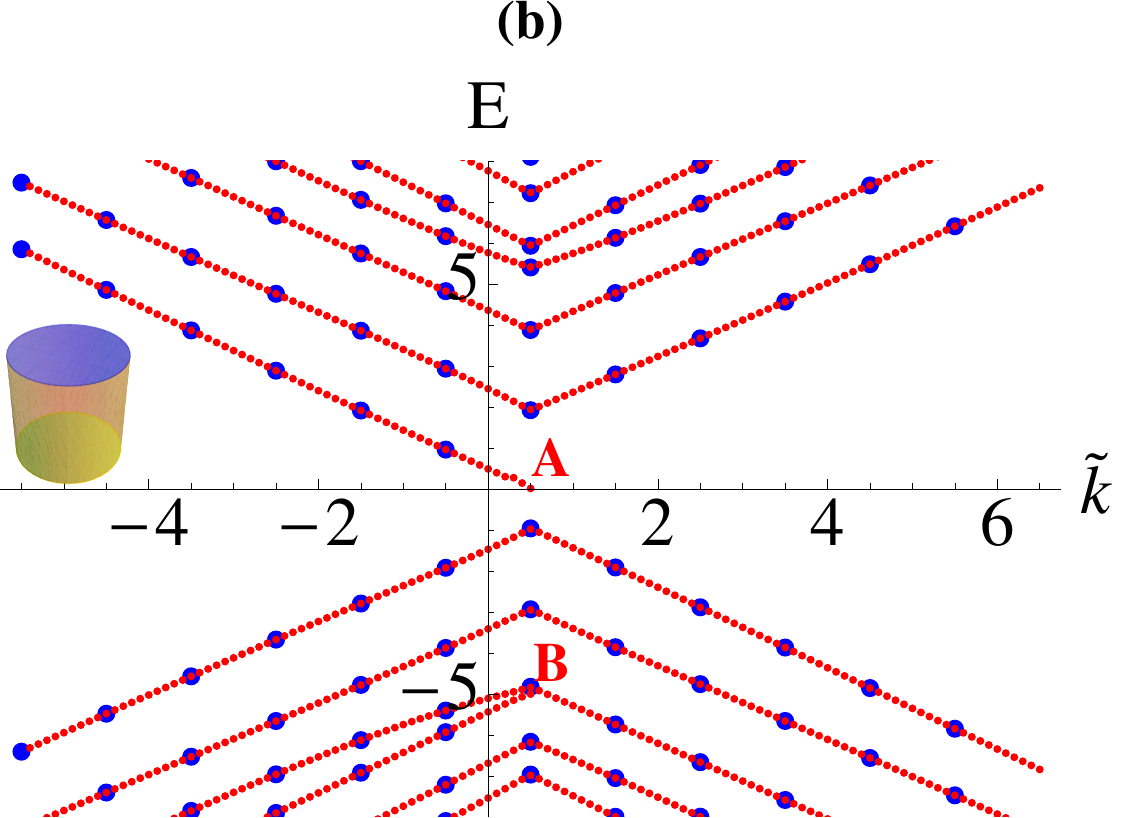}}
\quad
\subfigure{\includegraphics[width=.39\textwidth]{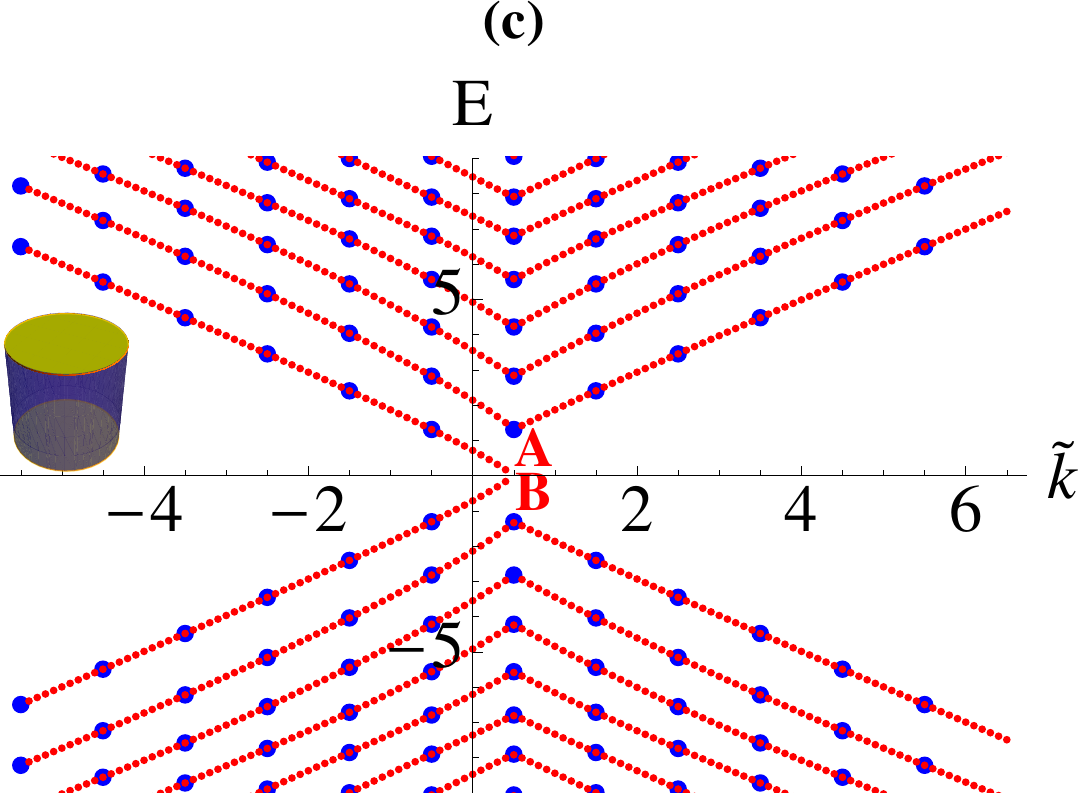}}
\quad
\subfigure{\includegraphics[width=.39\textwidth]{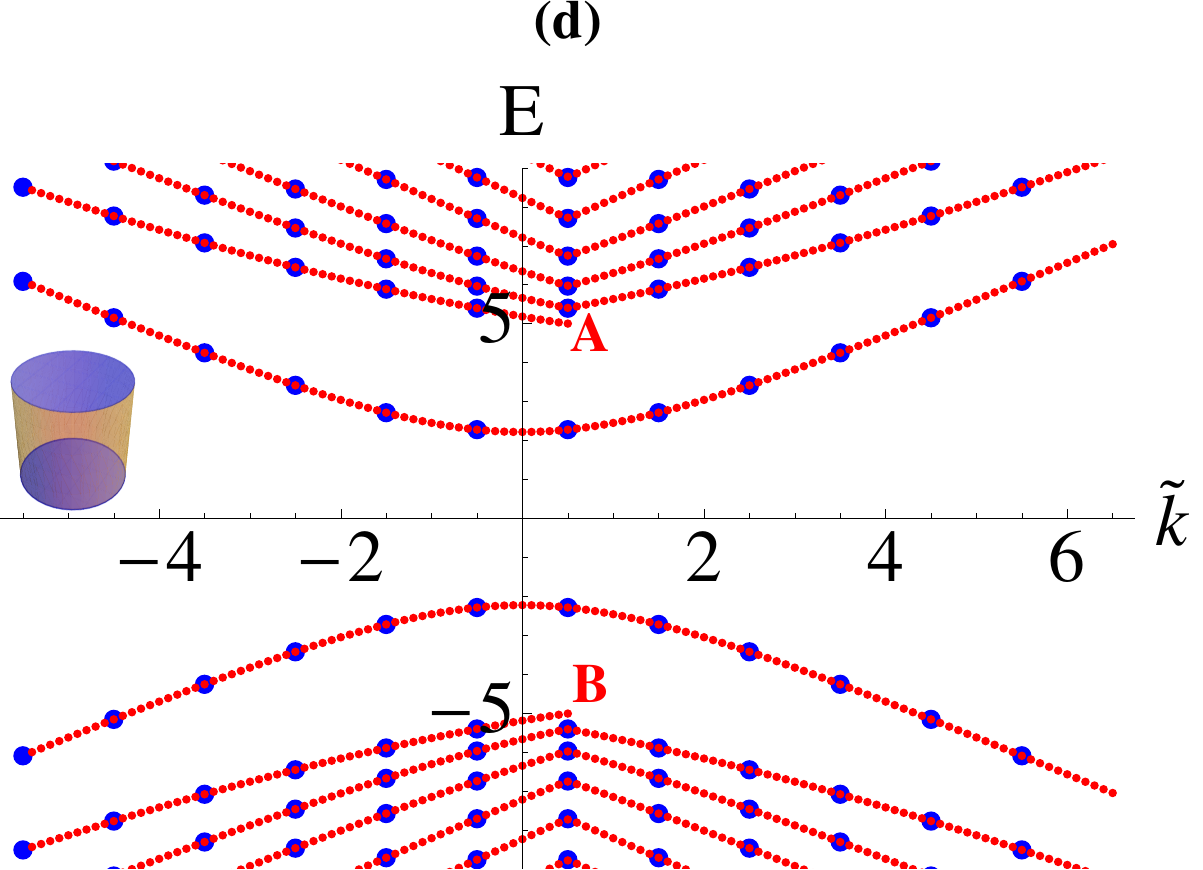}}
\quad
\subfigure{\includegraphics[width=.39\textwidth]{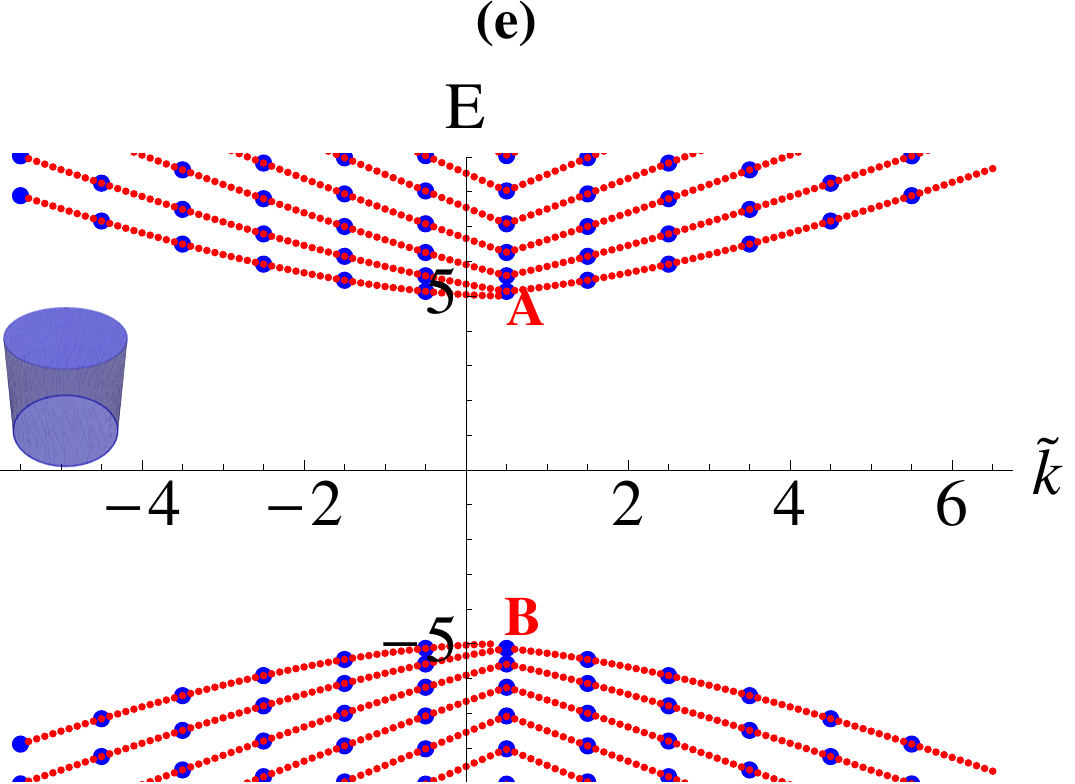}}
\quad
\subfigure{\includegraphics[width=.39\textwidth]{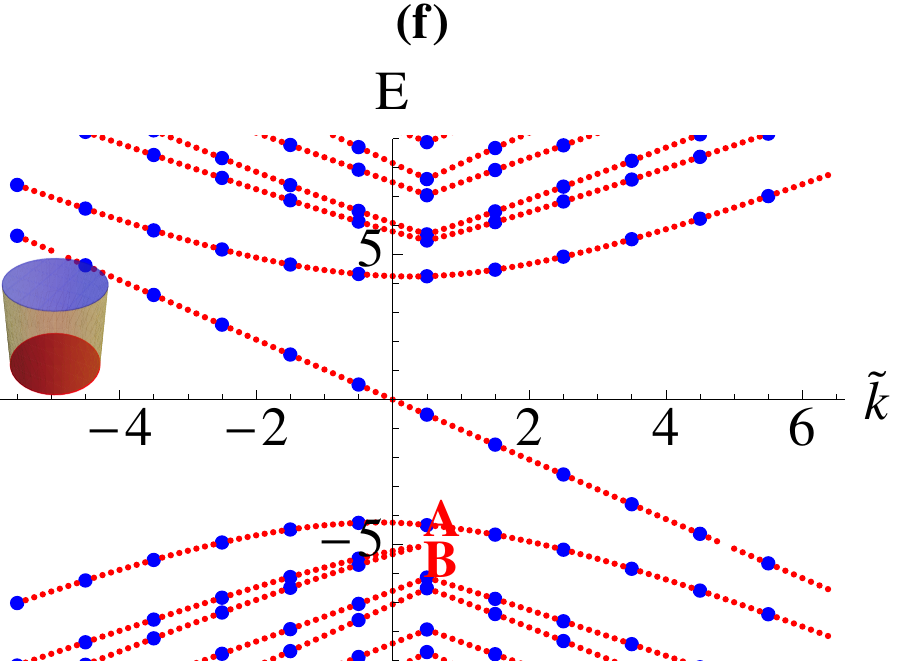}}
\caption{(Color online) The spectra corresponding to the cases (a) through (f) defined in the text. The insets reproduce cylinders for each case with color scheme: yellow for the massless regions, blue for positive mass, and in the case of (f), red for a negative orientated mass. Blue (larger) points at half-integer azimuthal number $\tilde{k}$ are the spectra with no flux inserted. The red (smaller) points represent the spectral flow as a localized flux is inserted as described in the text and shown in figure~\ref{cyl}. After one unit of flux, new states appear at marked points A and B. For all cases $|m|R=5$, $d/R=1/2$, and the energy, $E$, is in units of $1/R$. \label{cylres}}
\end{figure}

Figure \ref{cylres} shows the spectrum for the different cases as a function of azimuthal quantum number $\tilde{k}$. As before, the large(blue) points represent the spectra with no flux, and the smaller(red) points are the evolution of the spectra as one unit of flux is inserted.  The effect of the flux is manifested by the appearance of extra states at the end of the cycle inserting one unit of flux (marked A and B). The creation of new states relative to the original vacuum will have a half charge associated with them (see~\ref{sec:anom}). For all cases a simple pattern emerges: ‘anomalous’ bands create extra states near $-m_{\I}$ and $m_{\III}$ (A and B points). One can show that the states are never exactly at those values, including $E=0$ for the massless case. All other bands flow back to, or very near, the original spectrum. The difference between the cases occurs in the relative distribution of the wavefunctions after one unit of flux is inserted. This is now described. The Fermi energy is assumed to be at zero or mid-gap.

\begin{figure}
\centering
\subfigure{\includegraphics[width=.4\textwidth]{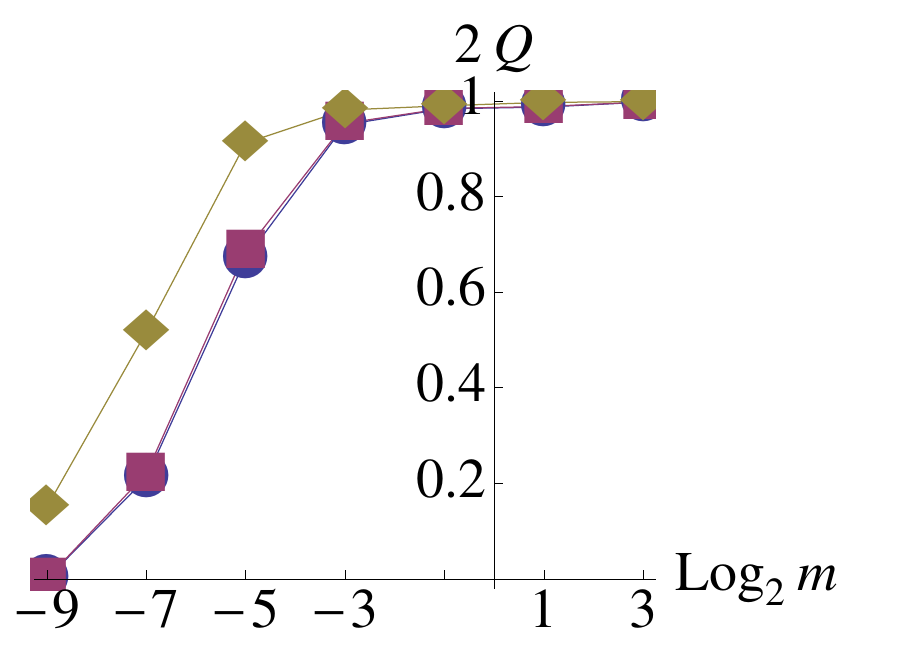}}
\subfigure{\includegraphics[width=.4\textwidth]{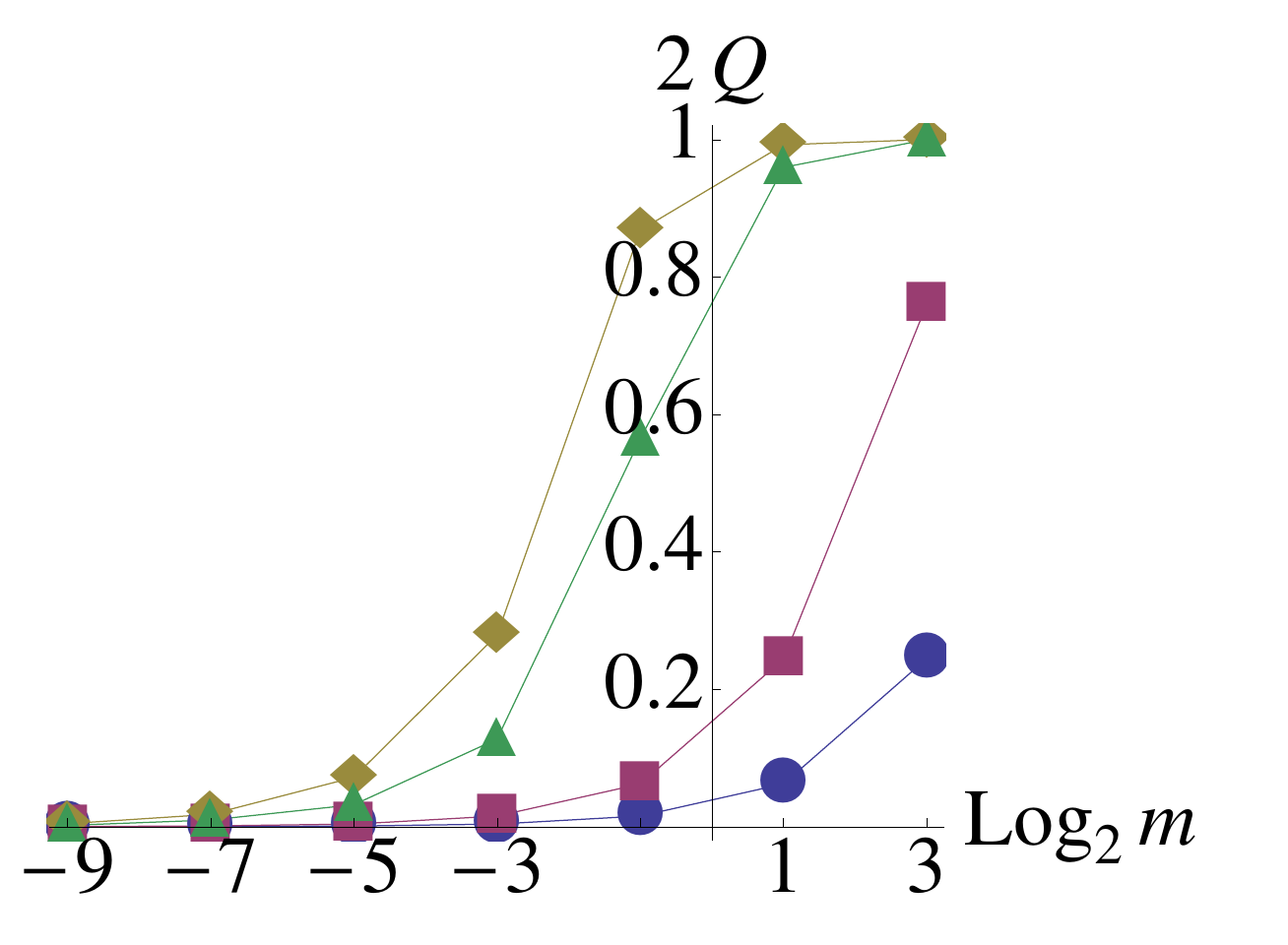}}
\\
\subfigure{\includegraphics[width=.4\textwidth]{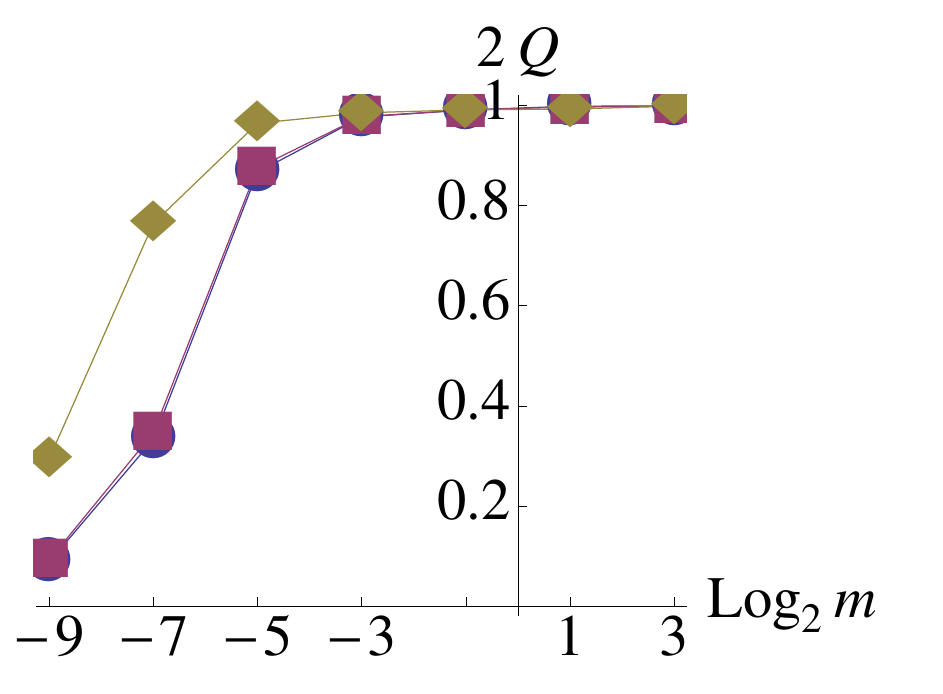}}
\subfigure{\includegraphics[width=.4\textwidth]{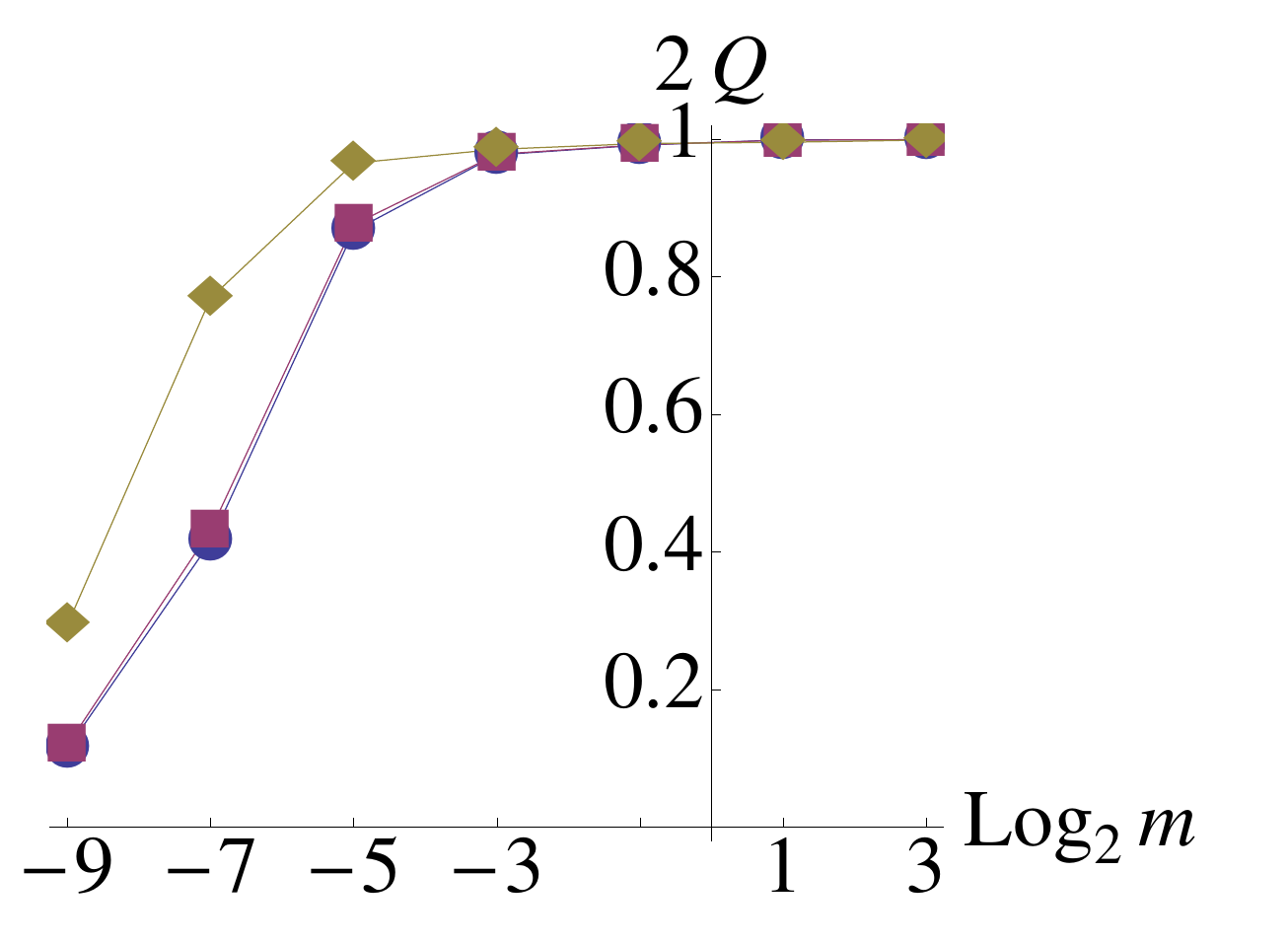}}
\caption{(Color online) The total charge $\times2$ (Q in units of $-e/2$) in the top cap after one unit of flux is inserted; cases (b) through (e) clockwise, with (b) starting at top-left. The horizontal axis plots a log scale of $m$ which is physically $mR$ since units are normalized by $R$. The various lines are for ratios $d/R=1/32, 1/8, 2$ for the circle(blue), square(purple), and diamond(yellow) respectively. The side-mass case also has $d/R=1$ shown in triangle(green). \label{chargepump}}
\end{figure}

\emph{Partial quantization.} First, in case (a) (massless case) no charge accumulation is seen. Indeed, the two states that appear near $E=0$ \emph{each} are equally split between the top and bottom flux-piercing, so that no net charge density appears accumulated.  In cases (b) through (e), the new states are now unevenly localized resulting in a net charge pumped. The negative energy states are localized near the bottom cap flux while the positive near the top. However, the total charge pumped to(from) the cap depends on the mass-region strength, going to zero as this value goes to zero. Quantitative results for the pumped charge are shown in figure~\ref{chargepump}, showing incomplete quantization if the mass-region is too small.

\emph{Extended Proximity Effect.} While the positive and negative bands both approach the value of the local mass, the net charge pumped responds to the configuration globally. A striking example is case (c), where charge is pumped to the caps, despite the mass being only present on the side and the similarity of the spectrum with case (a). Also in case (b), even though the mass in present on one cap, the amount of charge pumped is equal and symmetric on the top and bottom caps. In case (b) the equality of $|Q_{top}|=|Q_{bot}|$ is verified to high accuracy for all the masses shown in figure~\ref{chargepump} up to long cylinders of ratio $d/R=8$, thereby indicating that the whole cylinder behaves in a similar fashion as though a mass were present everywhere.

\emph{Opposite phases and chiral mode.} An example of the expected chiral band separating the two Hall phases is shown in this case (f). As in the previous cases two new states now appear at $E=-|m|$. Both of these states represent a half-charge at each cap relative to the flux-less vacuum. After unit flux these states give a deficit from each cap while a chiral mode becomes occupied.

As a final note, if the solenoid could pierce only one cap (a net outward flux),then a single $E=\pm m$ (with $m$ the mass of that cap with flux) state would be found for $\Phi=1$ in agreement with Index Theorems (see references within~\cite{Jackiw:1984ji, PhysRevD.16.1052}). In the present case the pair of new states ensure the overall system remains neutral but the relative distribution of charge does not remain uniform.

\subsection{The wormhole effect and surface theory fidelity}
\label{sec:closeda}
Before concluding with the examples, I wish to comment on whether the closed cylinder results of the previous section are applicable for a real system. It was argued in reference~\cite{Rosenberg} that surface electrons will tunnel through the bulk of a topological insulator for a very localized flux or if a hole is bored through the material. This was called the wormhole effect. Nevertheless, I will consider this question open for a general flux. Noting the results of section~\ref{sec:torus} one notices similarities with the work of reference~\cite{Rosenberg}. While I considered a smooth case, with a finite-size gap $1/L_2$ (blue points of figure~\ref{torusres}) it is clear such a gap will be of order $1/R$ where $R$ is the radius of the interior of a bored cylinder. In these cases, the effect of inserting a flux tube through a bored hole or a single plaquette of the microscopic lattice, can be removed by simply shifting the azimuthal number and the spectral flow will be similar to figure~\ref{torusres}(a). 

\begin{figure}
\centering
\subfigure[Microscopically localized flux]{
\raisebox{-.15cm}{\includegraphics[width=.16\textwidth]{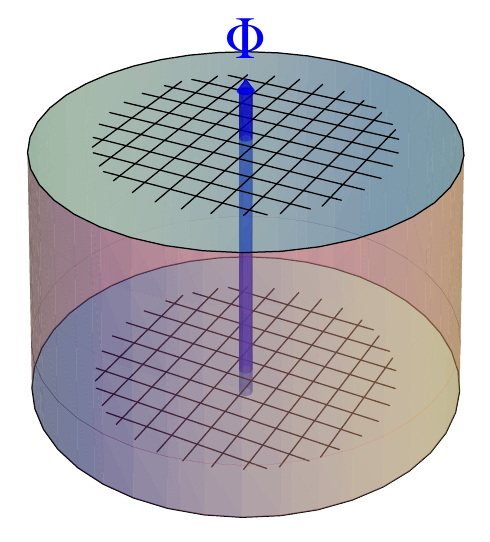}}
\raisebox{.27cm}{\includegraphics[width=.022\textwidth]{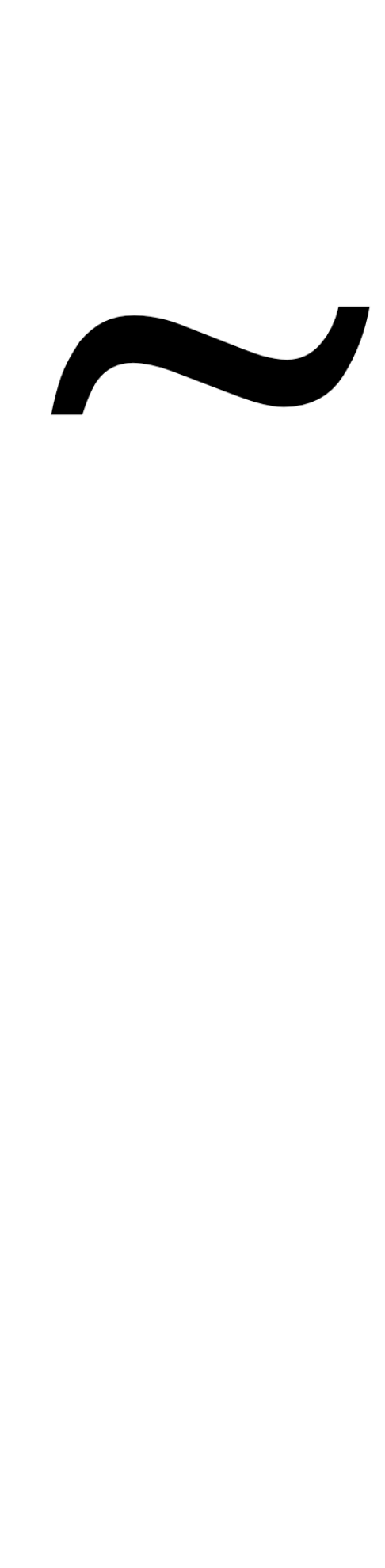}}
\raisebox{-.5cm}{\includegraphics[width=.22\textwidth]{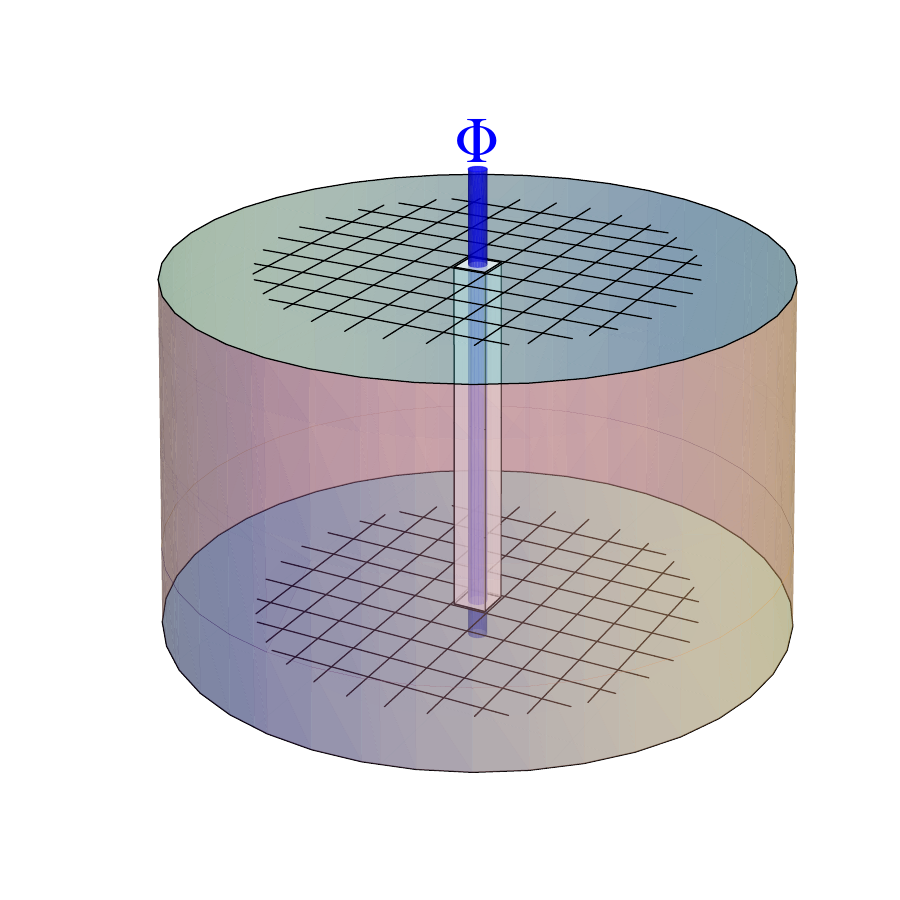}}
\raisebox{.27cm}{\includegraphics[width=.022\textwidth]{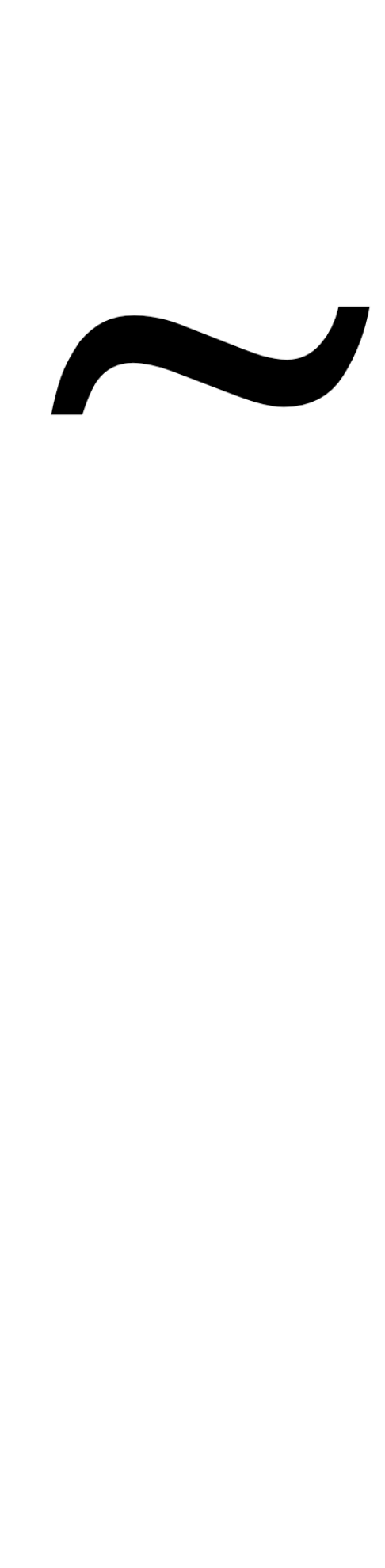}}
\includegraphics[width=.19\textwidth]{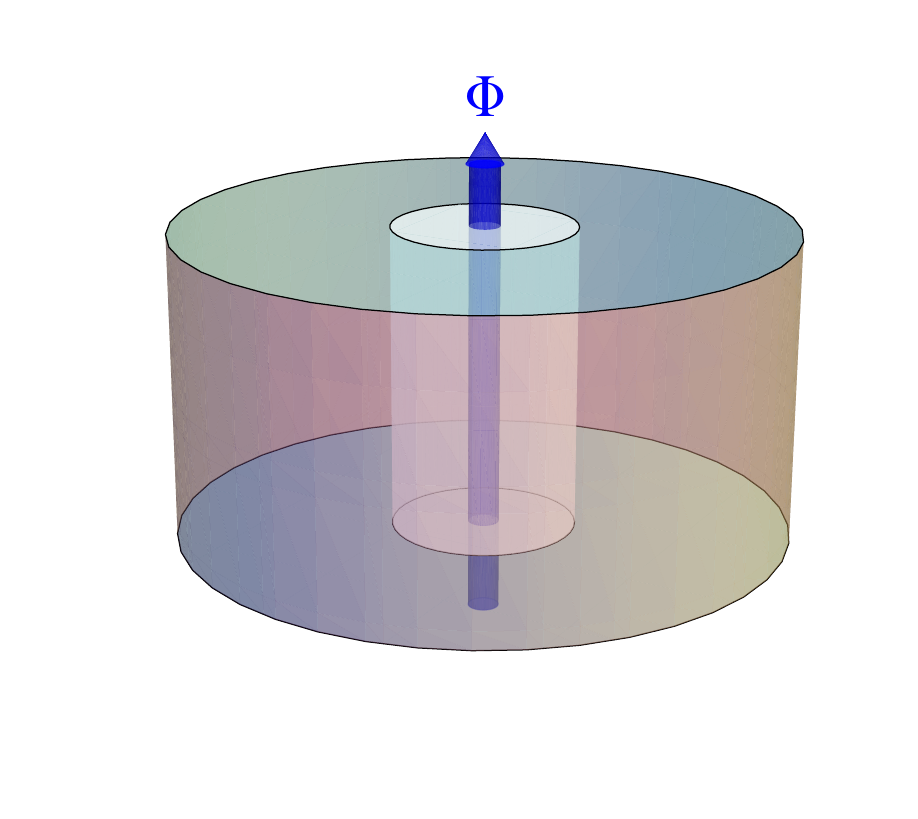}
}
\\
\subfigure[Microscopically diffuse flux]{
\includegraphics[width=4cm]{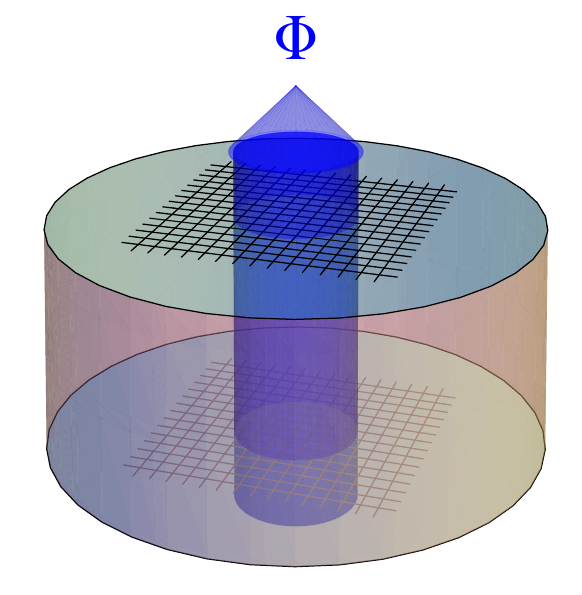}
\quad
\raisebox{.5cm}{\includegraphics[width=.029\textwidth]{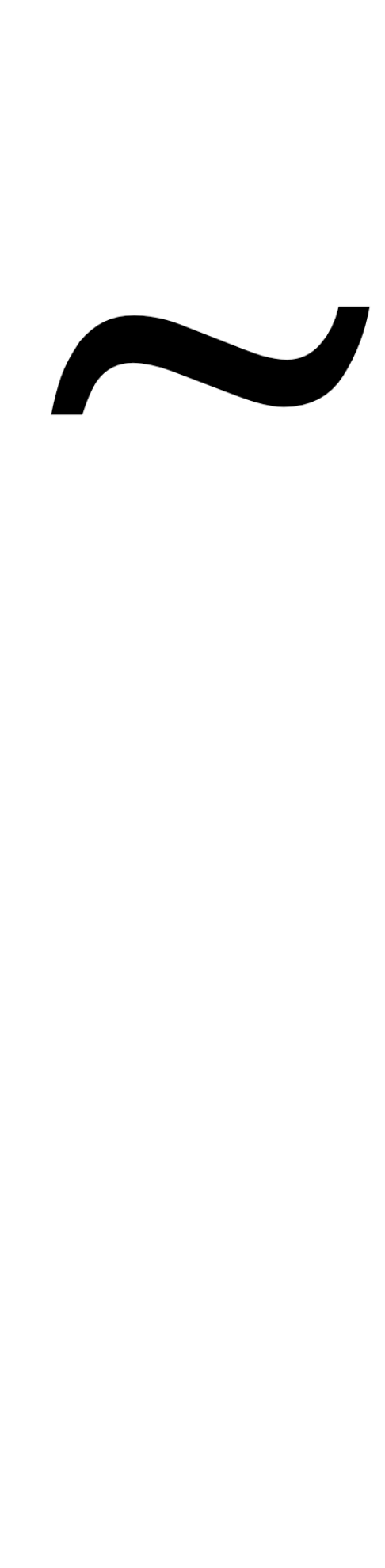}}
\quad
\raisebox{.25cm}{\includegraphics[width=.22\textwidth]{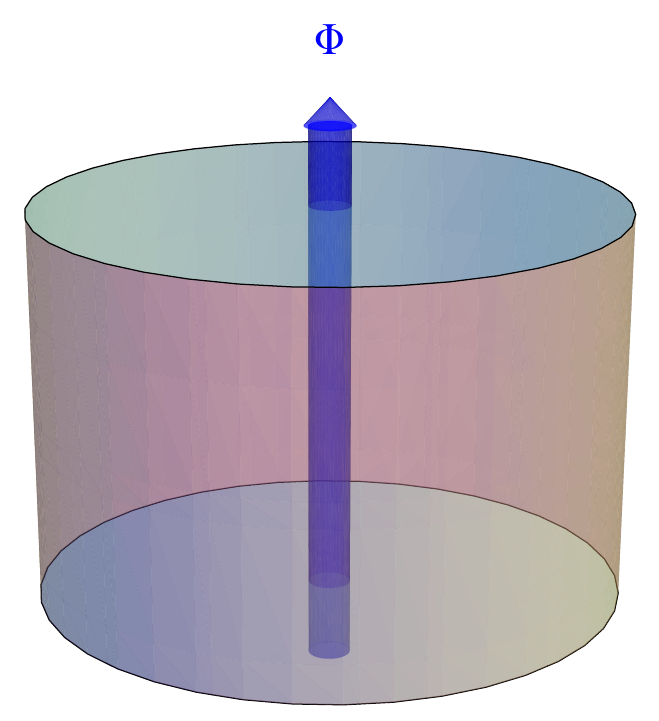}}
}
\caption{(Color online) Illustration contrasting a microscopically localized flux (a), which is qualitatively equivalent to the genus one surface discussed in section~\ref{sec:torus}, and a diffuse flux (b) which remains faithfully described by a surface theory of genus zero and penetrating flux. \label{difBf}}
\end{figure}

In fact, a similar picture will result whenever a simple global transformation can remove the effect of a threaded flux even if it goes through a hole or single plaquette. Thus the following ansatz is proposed: if an extremely localized flux string manages to pierce a single or a few plaquettes throughout the bulk, then currents can be described by a ``surface'' of genus one as in figure~\ref{difBf}(a), since either for a lattice or continuum theory the flux simply shifts the azimuthal number.
The spectrum of the surface theory of the torus contains the gap-closing seen in reference~\cite{Rosenberg} and corresponds to the ability for electrons to propagate through the interior surface of the torus. As noted in section~\ref{sec:torus}, this band is actually not responsible for the anomalous current. In any case, the fact that charge will not accumulate with a mass partially covering the surface, is just a case of the extended proximity effect that forces the Hall current not to exhibit an edge.  

If instead the flux is smoothly varying and extended over many plaquettes, then it is still reasonable to expect the low energy theory on the surface be described faithfully by the Dirac theory but on a surface of genus zero (this is illustrated in figure~\ref{difBf}(b)). In this case it turns out that the physical inclusion of the flux in the manifold matters beyond simply shifting the azimuthal angle, and a fractional localized charge might be observed. 

This potential cross-over is supported by the results of reference~\cite{PhysRevLett.105.190404} which numerically solves for the case of a constant magnetic field through the two disjoint surfaces: a cube with x and y faces identified and open z faces. They find the expected fractional charge smoothed out over each face, is preserved.

%%%%%%%%%%%%%%%%%%%%%%%%%%%%%%%%%%%%%%%%%%%%%%%%%%%%%%%%%%%%%%%%%%%%%%%%%%%%
%%%%%%%%%%%%%%%%%%%%%%%%%%%%%%%%%%%%%%%%%%%%%%%%%%%%%%%%%%%%%%%%%%%%%%%%%%%%
\section{Flux-Charge Qubit}
\label{sec:qubit}

\begin{figure}
\centering
\includegraphics[width=.6\textwidth]{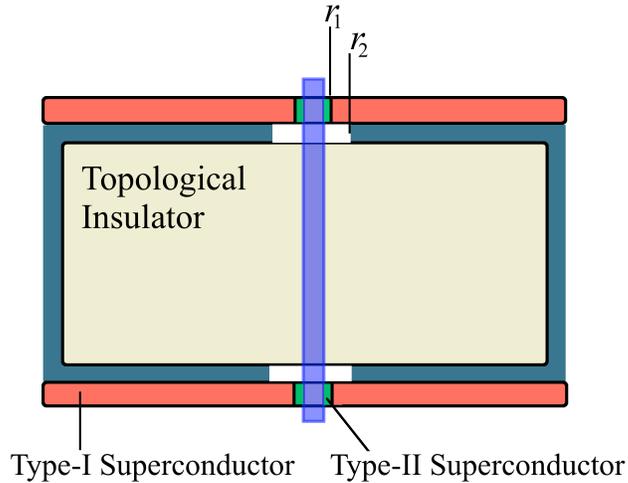}
\caption{(Color online) Cross-section of a configuration supporting a qubit. The blue-green band surrounding the topological insulator represents a ferromagnetic mass. See text for detailed explanation. \label{qubit}}
\end{figure}

The pattern of the results from the previous section implies that unit fluxes induce new states with energy equal to the local mass (as opposed to the Hall current which depends on the masses globally). Taking this into account, it is possible to conceive of a configuration that will produce close to degenerate $E=0$ states separated by an arbitrary energy gap--a desirable property for a qubit.  Qubits using topological insulators have been proposed in reference~\cite{PhysRevLett.100.096407} involving Majorana fermions. The current proposal would use (fractional) electric charge to distinguish between the states, while requiring magnetic flux for their stabilization. It might therefore be considered a `flux-charge' qubit. The basic idea is shown in figure~\ref{qubit}, with details described below. One imagines the cylinder coated with mass almost entirely (shown in blue around the topological insulator surface) except for a region of radius $r_2$. If $r_2\sim 1/m$ then any state localized in $r_2$ will be gapped by order $m$ as well, thereby making the entire spectrum gapped by $m$. Now if a localized flux can be inserted within $r_2$, with width of say $r_1$, it will induce a low energy states which in the limit of an ultra-localized flux will approach $E=0$ (two such states localized on the top and bottom windows as shown in figure~\ref{qubiteigen}).  

\begin{figure}
\centering
\includegraphics[width=.45\textwidth]{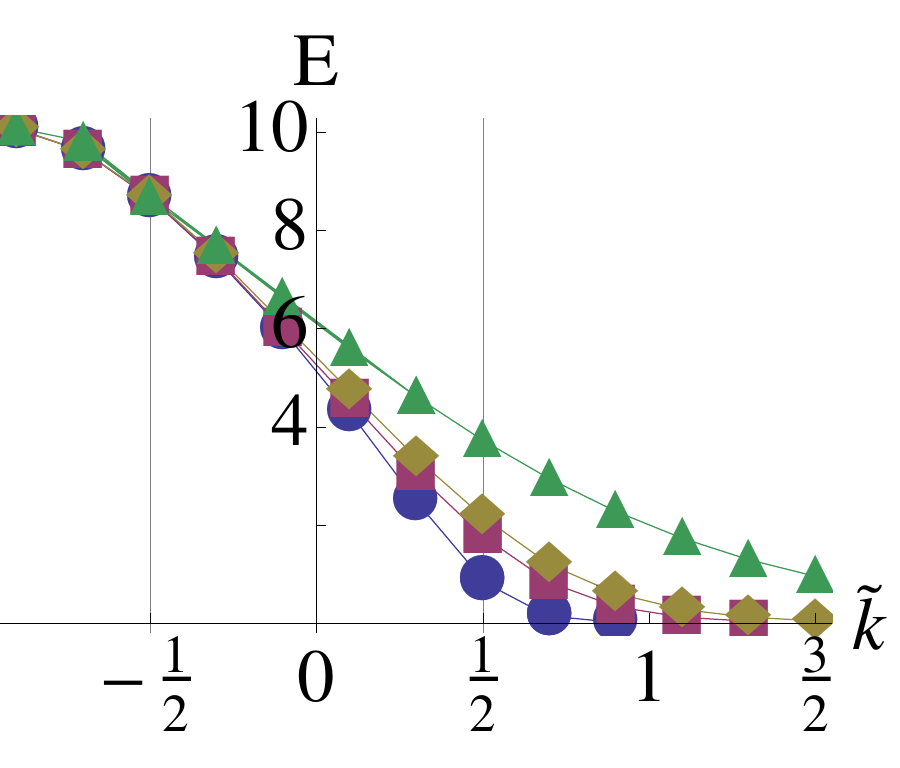}
\caption{(Color Online) The energies of the induced state as a localized flux is inserted for the configuration of figure~\ref{qubit}. The energy flow for up to two units of flux are shown. The four curves from top to bottom are for different flux radii: $r_1 = 0.5r_2, 0.1r_2, 0.05r_2, 0.001r_2$ respectively. All cases have $r_2= 0.1R$ and $m=10$ R$^{-1}$. States starting at $\tilde{k}= -1/2$ flow to $\tilde{k}= 3/2$, while a second state starting from $\tilde{k}=-3/2$ flows to $\tilde{k}=1/2$ along similar curves. Note that symmetric $E<0$ states and states that have energies $|E| \sim 10$ (as in figure~\ref{qubiteigen}) are not shown. See section~\ref{sec:discussion} for discussion on magnitudes. 
\label{qubitflow}}
\end{figure}

\begin{figure}
\centering
\includegraphics[width=.35\textwidth]{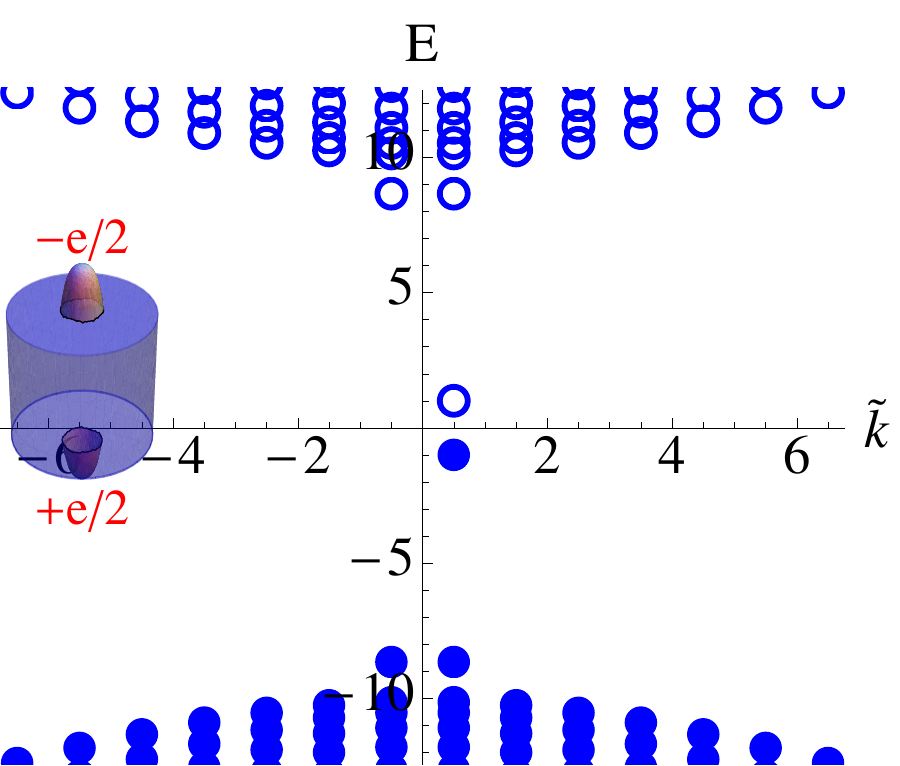}
\quad
\includegraphics[width=.35\textwidth]{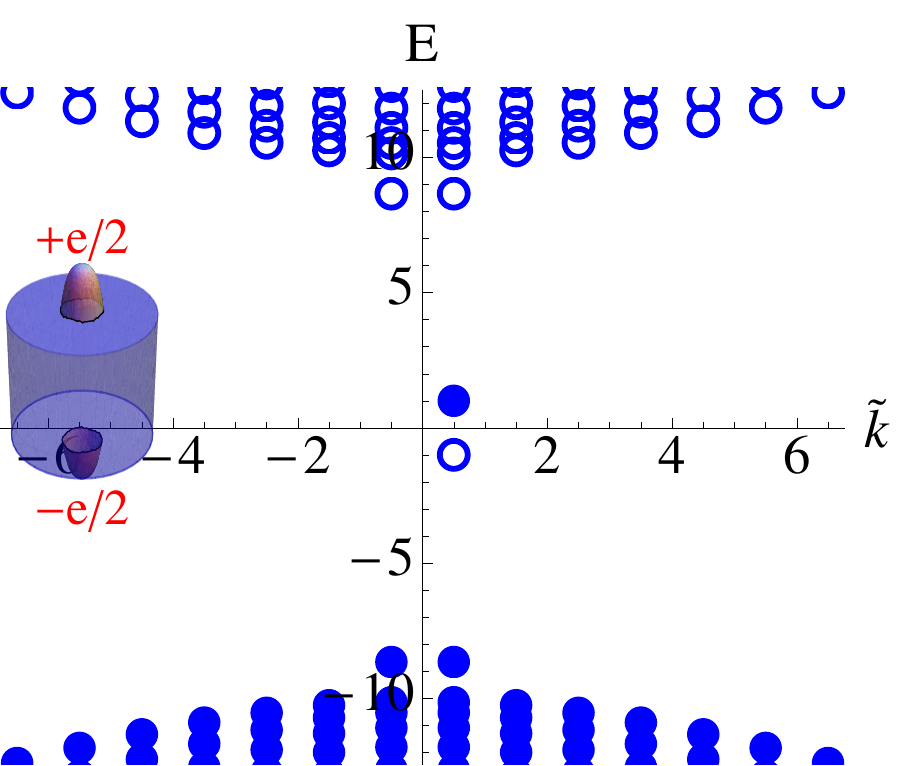}
\caption{(Color Online) Electron occupation (filled points) and insets showing charge density (plotted on the surfaces) for qubit states $|1\rangle$ (left) and $|2\rangle$ (right). Solutions are for the same parameters as those in figure~\ref{qubitflow}. \label{qubiteigen}}
\end{figure}

While in previous sections I have been considering a single unit of flux, when more than one unit of flux is inserted, other states which start at lower $\tilde{k}$ flow down into the gap towards E=0. The flow is along the same, or similar, trajectory. A generic picture is that of figure~\ref{qubitflow} which shows the flow of energy states as up to two units of flux are inserted, after which there will be a new pair of states with $E \sim \pm 0$ at $\tilde{k}=3/2$, and $\tilde{k}=1/2$, while the rest of the spectrum remains at $|E|\geq\; \sim m$. Each of these states will be localized in different regions: $E\sim+0$ states will be localized in the bottom flux-piercing while $E\sim -0$ states will be on top. Therefore $E>0$ and $E<0$ states will not be mixed by local noise (different $\tilde{k}$ states however can be). If, however, the flux is fractional there will in general be one state in the mid-gap region spoiling the energy separation (between $\tilde{k} = -1/2$ and $1/2$ in figure~\ref{qubitflow}). Therefore a limiting factor is the need for integer flux. A second issue, is that for the energies to flow to a value sufficiently close to $E=0$, $r_1$, the flux width, must be sufficiently small compared to $r_2$, otherwise the mass exerts a local proximity effect for the $\tilde{k}= 1/2$ state. From figure~\ref{qubitflow}, at least $r_1 < 0.1r_2$ is required for reasonable energy scale separation.


With interactions turned on, the true ground state is the most neutral occupation, which is zero for even flux and $\pm e/2$ for odd (split on caps). A fluctuation through an even flux would destroy the charge and so for all practical purposes precision up to a single flux is needed, or specifically exactly one flux quantum. 

These constraints may have a potential solution. To shield flux from regions where it is not wanted, a (type-I) superconductor can be put over the top and bottom surfaces as shown (figure~\ref{qubit}) with holes of radius $r_1$. The induced current of the superconductor would favor integer flux through $r_1$, which because it is small could easily be accommodated by a reasonable macroscopic magnetic field over the whole cap. To further stabilize the single flux constraint, a Type-II superconductor could be used with the temperature of the system tuned so that the vortex coherence length $\sim r_1$ favoring a single vortex. Other possibilities perhaps using SQUIDs might also be suited.

If a single flux quantum can be stabilized then the basis of the qubit are simply the occupation of $E \sim 0^{(+)-}$ state, as shown in figure~\ref{qubiteigen}, $|1\rangle = a^{\dagger}_{0^{-}}\left(\prod_{E\leq m}b_E^{\dag} \right)|0\rangle $, $|2\rangle = a^{\dagger}_{0^{+}}\left(\prod_{E\leq m}b_E^{\dag} \right)|0\rangle $ (in the convention of~\ref{sec:anom}). $|1\rangle$ has a $-e/2$ charge at the top window ($+e/2$ at bottom) and vice versa for $|2\rangle$. Then in this basis, applying an electric field, or a potential $V$ with $V|_{top}=-V_{bottom}=v$, generates a term in the Hamiltonian $\sim v\sigma^3$ in the qubit subspace. A $\sigma^1$ or $\sigma^{2}$ matrix element necessary for full unitary evolution, is less trivial. Effectively switching the electron occupation is required.
One potential route is simply inserting metallic electrodes connecting the top and bottom windows, allowing the electron to tunnel and with the aid of a bias voltage. Exploring these possibilities are left for future work. %The technicalities of this route and other possibilities, however, are left for future work.

%%%%%%%%%%%%%%%%%%%%%%%%%%%%%%%%%%%%%%%%%%%%%%%%%%%%%%%%%%%%%%%%%%%%%%%%%%%%
%%%%%%%%%%%%%%%%%%%%%%%%%%%%%%%%%%%%%%%%%%%%%%%%%%%%%%%%%%%%%%%%%%%%%%%%%%%%
\section{Discussion}
\label{sec:discussion}

In summary the details of how anomalous currents manifest themselves on the surface of topological insulators has been presented as well as a potential application in the form of a flux-charge qubit. Most striking is the extended proximity effect, which is nonetheless rendered reasonable by a lack of quantization. This effect is manifested through the non-local nature of the wavefunctions, despite the spectrum responding to mass terms locally. The lack of quantization is more physical than the sharp Sign($m$) dependence found for the infinite plane. The existence of a chiral band which was previously known, is consistent with the result that oppositely orientated Hall phases are fully quantized, so that Chiral band receives or releases a whole charge. An effective unquantized current is restored in the sense that the chiral band can be diffuse.

It was noted for the torus that $m L/2\pi \sim 20$ was necessary for the onset of full quantization while for the closed cap geometry, from figure~\ref{chargepump}, onset occurs at $mR \sim 1/8$ for massive caps and $md\sim 8$ for mass on the sides of the cylinder (case c). For a typical Fermi velocity~\cite{hsieh-2008-452, xia-2009-5, Zhang_Liu_Qi_Dai_Fang_Zhang_2009} $v_f = 10^{6}$ m/s, setting $L$, $d$, or $R$ on order of 1 mm would correspond to $m= 8\times 10^{-2}$, $5\times 10^{-3}$, $8\times10^{-5}$ meV respectively, or for $L$, $d$, $R$ of order 1 $\mu$m, full quantization begins at $m= 80$, $5$ and $8\times 10^{-2}$ meV. These are well within experimentally observed time-reversal breaking surface mass terms of order meV~\cite{Chen06082010}. The wide range of values suggests that the effect is strongly geometry dependent, and as such these values should only be taken as rough guides.

The required magnitudes for the qubit proposal are more constraining. Using the value for $v_f$ above, for a gap of $3^\circ $k $\sim 0.3$ meV, a value of $r_2 =  2.5$ $\mu$m is required. Therefore $r_1 \lesssim 0.1$ $\mu$m, which is small. The value of $R$ and the aspect ratio is essentially unconstrained. 

Finally, whether fractional charges remains intact for a diffuse enough magnetic field remains an open question which can be investigated both numerically and experimentally. If this is the case, then the flux-charge qubit could be supported and offer an alternative approach to quantum computation.

\section{Acknowledgments}
This research was originally stimulated by discussions with Dung-Hai Lee. I also greatly benefited from many useful and stimulating discussions with J\"{o}rg Gramich.

%%%%%%%%%%%%%%%%%%%%%%%%%%%%%%%%%%%%%%%%%%%%%%%%%%%%%%%%%%%%%%%%%%%%%%%%%%%%
%%%%%%%%%%%%%%%%%%%%%%%%%%%%%%%%%%%%%%%%%%%%%%%%%%%%%%%%%%%%%%%%%%%%%%%%%%%%
\appendix
\section{Spectral asymmetry and anomaly}
\label{sec:anom}
The quantum anomaly can be understood by two main routes. For a (2+1) dimensional Dirac theory on the infinite plane coupled to a background gauge field $A^{\mu}(x)$, one can simply compute the effective action and extract the current term \cite{Niemi:1983rq, Jackiw:1984ji}
\begin{eqnarray}
\label{anom}
\langle j^{\mu} \rangle  = i\frac{\delta}{\delta A_{\mu}}  \ln \det(i\slashed{\partial}+ e\slashed{A}-m) \\
 = \frac{e^2}{8\pi}\frac{m}{|m|}\epsilon^{\mu \nu \lambda}F_{\nu \lambda} +\dots \\
\therefore \; \langle Q \rangle  = \int \langle j^{0} \rangle = \frac{e}{2}\Phi
\end{eqnarray}
where the field strength $F_{\mu \nu}= \partial_{\mu}A_{\nu}- \partial_{\nu}A_{\mu}$ and $\Phi =-e\int d^2rB/(2\pi)$. While this gives an exact analytic result, explicit computation of the log of the determinant is difficult in all but the simplest cases such as the infinite plane. 

Alternatively one can diagonalize the single particle Dirac Hamiltonian $H$, $H\psi_{\lambda}(\vec{x})= E_{\lambda}\psi_{\lambda}(\vec{x})$. $\psi$'s are $n$-component classical spinors, $H$ contains $\gamma^\mu$ matrices satisfying $\{ \gamma^{\mu}(x) ,\gamma^{\nu}(x)\} = g^{\mu\nu}(x)$ and $g$ is the coordinate metric (see references~\cite{0201503972, Spinors}). Then the fermion annihilation operator at $\vec{x}$ can be expanded in this basis~\cite{PhysRevD.13.3398}:
\begin{equation}
\Psi(\vec{x})= \sum\limits_{o} a_{o}\psi_{o} +  \sum\limits_{+\lambda}   a_{+\lambda}\psi_{+\lambda}(\vec{x}) +  \sum \limits_{-\lambda}  b^{\dag}_{-\lambda}\psi_{-\lambda}(\vec{x})
\end{equation}
where $\psi_{\pm\lambda}$ are the positive(negative) energy eigenstates, $\psi_{o}$ are possible zero energy states, and $a, b$ are fermionic annihilation operators satisfying the usual commutation relations (here and below, $\lambda$ with no $\pm$ qualifier is taken to mean \emph{any} eigenstate).

Then the vacuum expectation of the current, $J^{\mu}(\vec{x}) =\frac{-e}{2}[ \Psi^{\dag}(\vec{x})\gamma^{0}\gamma^{\mu}\Psi(\vec{x})-
(\Psi(\vec{x}))^{Ts}(\Psi^{\dag}(\vec{x})\gamma^0\gamma^{\mu})^{Ts} ] $,  (with $\gamma=\gamma(\vec{x})$ and $\psi=\psi(\vec{x})$) can be written~\cite{PhysRevD.13.3398}:
\begin{eqnarray}
\label{currenti}
\frac{1}{-e}\langle J^{\mu}(\vec{x}) \rangle =&  -\frac{1}{2}\sum \limits_{\pm \lambda}\left( \psi^{\dag}_{+\lambda}\gamma^{0}\gamma^{\mu}\psi_{+\lambda}-
\psi^{\dag}_{-\lambda}\gamma^{0}\gamma^{\mu}\psi_{-\lambda}\right) \nonumber \\
 &+\frac{1}{2}\sum \limits_{o}\psi^{\dag}_{o}\gamma^{0}\gamma^{\mu}\psi_{o}-\frac{1}{2}\sum \limits_{u}\psi^{\dag}_{u}\gamma^{0}\gamma^{\mu}\psi_{u}
\end{eqnarray}
where $\pm \lambda$ refers to non-zero states and $o$, $u$ sum over occupied(unoccupied) zero modes. To extract the charge in one region versus another $\langle J^{0}(\vec{x})\rangle$ may be integrated over regions of interest.

\section{Wavefunctions and quantization condition on the torus}
\label{sec:sol_torus}
Although it is simple enough to write the Dirac equation and the form of solutions for both of the cases in the text, for clarity in notation, I will split them. For a review on vielbeins and spinors in curved space see references~\cite{Nakahara:2003, Eguchi:1980jx, Spinors}.

\subsection{Case 1: Fully massive torus; step potential in $\hat{\phi}$ direction}
As discussed in the text (with unit definitions), for a fully massive torus with a potential $V_{\alpha}=(-v, +v, -v)$ for regions I, II, III as in figure~\ref{toruscoords1} the Dirac equation is:
\begin{equation}
(-i\sigma^1\partial_{z}-i\sigma^2\partial_{\phi}+\sigma^3m
)\tilde{\psi}_{\alpha} =(E-V_{\alpha})\tilde{\psi}_{\alpha}
\end{equation}
To simplify the formulae I have put $v$ instead of $-ev/2$. In this case the $\hat{z}$ direction is trivial and solutions are of the form $\tilde{\psi}=\exp(i k z)f_{\alpha}(\phi)$ with $k \in \mathbb{Z}$ (the same $k$ in all regions). If $E\neq \pm \sqrt{k^2+m^2} \pm v$ and $E\neq \pm |k|$, $f_{\alpha}(\phi)$ is given by:
\begin{eqnarray}
f(\phi)=&A_{\alpha}  \left( \begin{array}{c}1 \\ 
\frac{i\sqrt{(E-V_{\alpha})^2-k^2-m^2}+k}{E-V_{\alpha}+m}\end{array} \right)  e^{i\sqrt{(E-V_{\alpha})^2-k^2-m^2}\phi}
\nonumber \\
&+ B_{\alpha} \left( \begin{array}{c} 1\\ 
\frac{-i\sqrt{(E-V_{\alpha})^2-k^2-m^2}+k}{E-V_{\alpha}+m}\end{array} \right) e^{-i\sqrt{(E-V_{\alpha})^2-k^2-m^2}\phi}
\end{eqnarray}
The coefficients in each region and energy quantization are determined by normalization and matching conditions $f_{\I}(L_{2}/2)=\eta f_{\III}(-L_{2}/2)$, $f_{\I}(l/2)=f_{\II}(l/2)$ and $f_{\III}(-l/2)=f_{\II}(-l/2)$ with $\eta=-1$ for the chosen spin-structure in the $\hat{\phi}$ direction. The quantization condition is:
\begin{equation}
|a|^2(\cos(pl+q(L_{2}-l))-1)=
|b|^2(\cos(pl-q(L_{2}-l))-1)
\end{equation}
where I have defined $q= \sqrt{(E+v)^2-k^2-m^2}$, $p= \sqrt{(E-v)^2-k^2-m^2}$, $a=\frac{k+ip}{E-v+m}-\frac{k-iq}{E+v}$,
and $b=\frac{k+ip}{E-v+m}-\frac{k+iq}{E+v}$. This is solved numerically and shown (for a representative case) in figure~\ref{torusEres}. The special cases $E\neq \pm \sqrt{k^2+m^2} \pm v$ and $E\neq \pm |k|$ (and special sub-cases of these such as $E=0$) must be treated separately although in a similar manner as above. The solutions in those cases are linear in one of the regions, and in general do not yield new solutions.

\subsection{Case 2: Partially massive strip}
In this case the Dirac equation is:
\begin{equation}
(-i\sigma^1\partial_{z}-i\sigma^2\partial_{\phi}+\sigma^3m_{\alpha}
)\tilde{\psi}_{\alpha} =E\tilde{\psi}_{\alpha}
\end{equation}
where $m_{\alpha}=(0, m, 0)$ for regions I, II, III as in figure~\ref{toruscoords2}.
Different spin-structures are encoded in whether the boundary condition on $\tilde{\psi}$ is anti-periodic or periodic in $\phi$. The 
In each region I, II, III the solutions are of the form $\tilde{\psi}=\exp(i\tilde{k}\phi)f(z)$ with $\tilde{k}\in \mathbb{Z}+1/2$. If $E\neq \pm \sqrt{\tilde{k}^2+m^2}$ and $E\neq \pm |\tilde{k}|$, $f(z)$ is given by:
\begin{eqnarray}
f(z)=&A\left( \begin{array}{c}   1\\ 
\frac{\sqrt{E^2-\tilde{k}^2-m^2}+i\tilde{k}}{E+m}\end{array}\right) e^{i\sqrt{E^2-\tilde{k}^2-m^2}z} \nonumber
\\ &+
B\left( \begin{array}{c} 1\\ 
\frac{-\sqrt{E^2-\tilde{k}^2-m^2}+i\tilde{k}}{E+m}\end{array} \right)e^{-i\sqrt{E^2-\tilde{k}^2-m^2}z}.
\end{eqnarray}
where the region label has been omitted and the mass $m$ is understood to be zero in region I and III. The coefficients in each region and energy quantization are determined by normalization and matching conditions $f_{\I}(L_2/2)=\eta f_{\III}(-L_2/2)$, $f_{\I}(l/2)=f_{\II}(l/2)$ and $f_{\III}(-l/2)=f_{\II}(-l/2)$ with $\eta=1$ for the chosen spin-structure along $\hat{z}$ loops. The quantization condition is as before:
\begin{equation}
|a|^2\cos(pl+q(L_2-l))=
|b|^2\cos(pl-q(L_2-l))
\end{equation}
but in this case the definitions are $q= \sqrt{E^2-\tilde{k}^2}$, $p= \sqrt{E^2-\tilde{k}^2-m^2}$, $a=\frac{\tilde{k}+ip}{E+m}-\frac{\tilde{k}-iq}{E}$,
and $b=\frac{\tilde{k}+ip}{E+m}-\frac{\tilde{k}+iq}{E}$. This is solved numerically and shown (for a representative case) in figure~\ref{torusres}. The special cases $E= \pm \sqrt{\tilde{k}^2+m^2}$ and $E= \pm |\tilde{k}|$ (and special sub-cases of these such as $E=0$) must be treated separately although in a similar manner as above. The solutions in those cases in general do not yield new solutions.

\section{Derivation of the Dirac equation on closed cylinder surface}
\label{sec:eqn_cyl_surf}
\begin{figure}
\centering
\includegraphics[width=.45\textwidth]{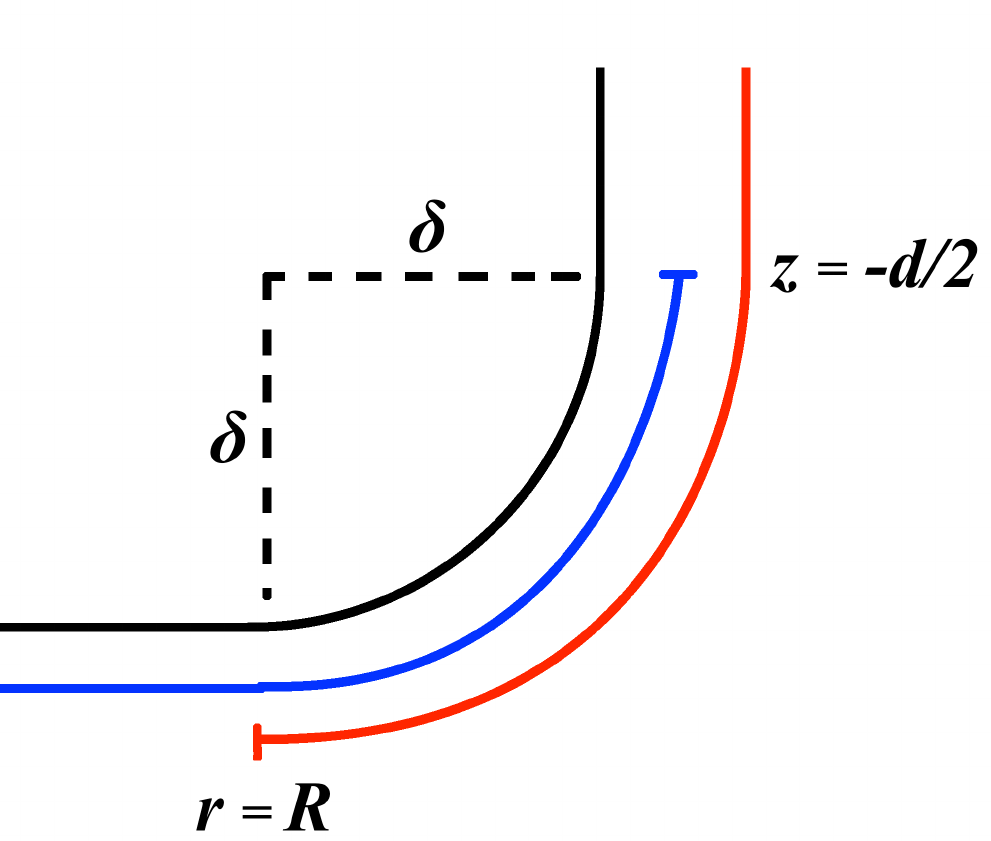}
\caption{(Color online) Side view of smoothed out cylinder edge. \label{corner}}
\end{figure}

In the notation of figure~\ref{cyl}, the closed cylinder is divided into three regions. The local form on the caps is as the Dirac equation for the disk described in equation~\ref{cart} which can be gotten by transforming the Dirac equation in Cartesian coordinates into polar coordinates. To guess the form on side, region II, one cannot simply replace $r\rightarrow R$ as the metric suggests. In particular the spin connection is wrong, and stems from the fact that all the curvature is at the sharp corner where the coordinate transformation is required. For a review on vielbeins and spinors in curved space see references~\cite{Nakahara:2003, Eguchi:1980jx, Spinors}. A real manifold must be smooth and the transition between charts should be a map between $\mathbb{R}^{2}\rightarrow\mathbb{R}^2$, although in practice one dimension can be made of infinitesimal width if the manifold is smooth. This suggest a more careful procedure is to smooth the corners into a semi-circle of radius $\delta$ as in figure~\ref{corner}. 
Replacing the relation $\delta^2=(r-R)^2+(z+\frac{d}{2})^2$ into the 3-dimensional Cartesian metric, the induced metric is, 
\begin{equation}
g= -\frac{\delta^2}{\delta^2-(r-R)^2}dr\otimes dr -r^2d\phi\otimes d\phi 
\end{equation}
and leads to the Dirac equation (using $\phi$ dependent vielbeins) in $(r, \phi)$:
\begin{eqnarray}
e^{-i\frac{\sigma^3}{2}\phi} 
\Bigg(  -i& \sigma^1 \frac{\sqrt{\delta^2-(r-R)^2}}{\delta}\partial_{r}-i\sigma^2\frac{\partial_{\phi}}{r}
\\
&-i\sigma^1\frac{\sqrt{\delta^2-(r-R)^2}}{2\delta r}
\Bigg) e^{i\frac{\sigma^3}{2}\phi}\psi =E\psi.
\end{eqnarray}
As $r\rightarrow R$, the equation becomes
\begin{eqnarray}
e^{-i\frac{\sigma^3}{2}\phi}
\Big( -i\sigma^1\partial_{r}-i\sigma^2\frac{\partial_{\phi}}{R}
-\frac{i\sigma^1}{2R}
\Big) e^{i\frac{\sigma^3}{2}\phi}\psi =E\psi,
\end{eqnarray}
just what we expect on the cap evaluated at $r=R$. However, replacing $r(z)$ and taking $z\rightarrow \frac{-d}{2}$ one gets:
\begin{eqnarray}
e^{-i\frac{\sigma^3}{2}\phi}
\Big( -i\sigma^1\partial_{z}-i\sigma^2\frac{\partial_{\phi}}{R+\delta}\Big) e^{i\frac{\sigma^3}{2}\phi}\psi =E\psi,
\end{eqnarray}
just what one expects from the side using an embedding, without any further transformation of $\psi$ i. e. $\tilde{\psi}_{\III}(R, \phi)= \tilde{\psi}_{\II}(-d/2, \phi)$. Less trivially is the boundary condition between the top and the side. The derivation proceeds in the same way if the local $\hat{z}$ were facing down. So if one defines a new $z_n=-z$, and calls the tentative wavefunction $\psi_{\II n}$ with the notation $\tilde{\psi}_{\II n}=e^{i\frac{\sigma^3}{2}\phi}\psi_{\II n}$ one would have found $(-i\sigma^1\partial_{z_n}-i\sigma^2\frac{\partial_{\phi}}{R})\tilde{\psi}_{\II n}(z_n, \phi) =E\tilde{\psi}_{\II n}(z_n, \phi)$ (with boundary condition $\tilde{\psi}_{\I}|_{r=R}=\tilde{\psi}_{\II n}|_{z=d/2}$) which is $(+i\sigma^1\partial_{z}-i\sigma^2\frac{\partial_{\phi}}{R})\tilde{\psi}_{\II n}(z, \phi) =E\tilde{\psi}_{\II n}(z, \phi)$. Evidently, then, the new $\tilde{\psi}_{\II n}= \sigma^2\tilde{\psi}_{\II}$, since $\sigma^2\sigma^1\sigma^2=-\sigma^1$. 

In summary equations~\ref{diraccyl1} and~\ref{diraccyl2} are obtained.

\section{Wavefunctions and quantization condition on cylinder}
The local solutions to equation~\ref{diraccyl1} (if there is no flux) are for the caps $\tilde{\psi}_{\I, \III}=\exp(i\tilde{k}\phi)f_{\I, \III}(r)$. If $E\neq \pm \sqrt{\tilde{k}^2+m_{\I, \III}^2}$ and $E\neq\pm m_{\I, \III}$, $f(r)$ has the form:
\begin{eqnarray}
\label{cap}
f(r)=&C_{1}\left( \begin{array}{c}  J_{\tilde{k}-\frac{1}{2}}(\sqrt{E^2-m^2}r)\\ \frac{i\sqrt{E^2-m^2}}{E+m}J_{\tilde{k}+\frac{1}{2}}(\sqrt{E^2-m^2}r)\end{array} \right)
\nonumber \\ &+
C_{2}\left( \begin{array}{c} Y_{\tilde{k}-\frac{1}{2}}(\sqrt{E^2-m^2}r)\\ \frac{i\sqrt{E^2-m^2}}{E+m}Y_{\tilde{k}+\frac{1}{2}}(\sqrt{E^2-m^2}r)\end{array} \right)
\end{eqnarray}
where a region label $=$I, III should be understood for the mass $m$, $f(r)$ and the coefficients $C_1$ and $C_2$. $\tilde{k} \in \mathbb{Z}+1/2$ is globally the same value for all regions, and different $\tilde{k}$ are linearly independent as required from the boundary condition along the $\phi$ direction (they are good quantum numbers). $J_{n}(r)$, $Y_{n}(r)$ are Bessel functions, the two independent solutions  satisfying $r^2 \frac{d^2 h}{dr^2} + r \frac{dh}{dr} + (r^2 - n^2)h = 0$, $J, Y= h(r)$ ($J_{n}(r)$ and $J_{-n}(r)$ also work if $n$ is not an integer). The solution on the side is $\tilde{\psi}_{\II}=\exp(i\tilde{k}\phi)f_{\II}(r)$, and if $E\neq \pm \sqrt{\tilde{k}^2+m_{\II}^2}$ and $E\neq\pm m_{\II}$, $f(r)$ is (again omitting region II label):
\begin{eqnarray}
f(z)=&A\left( \begin{array}{c}1\\ 
\frac{\sqrt{E^2-\tilde{k}^2-m^2}+i\tilde{k}}{E+m}\end{array} \right) e^{i\sqrt{E^2-\tilde{k}^2-m^2}z} \nonumber
\\& +
B \left( \begin{array}{c} 1\\ 
\frac{-\sqrt{E^2-\tilde{k}^2-m^2}+i\tilde{k}}{E+m}\end{array} \right)e^{-i\sqrt{E^2-\tilde{k}^2-m^2}z}.
\end{eqnarray}

Square-integrability at the origin sets one of the coefficients of the cap to zero, when $|\tilde{k}|\geq 1/2$. Using the boundary conditions discussed in the previous section, the remaining coefficients are determined along with the quantization condition:
\begin{equation}
\label{cylquant}
e^{-2d\sqrt{E^2-\tilde{k}^2-m^2_{II}}}\left(\frac{ih_{III}\gamma^*-g_{III}}{g_{III}+ih_{III}\gamma}\right)
=\frac{ig_{I}\gamma^*-h_{I}}{h_{I}+ig_{I}\gamma}
\end{equation}
with $h_{\alpha}= \frac{\sqrt{E^2-m^2_{\alpha}}}{E+m_{\alpha}}J_{\tilde{k}+\frac{1}{2}}(\sqrt{E^2-m^2_{\alpha}}R)$,
$g_{\alpha}= J_{\tilde{k}-\frac{1}{2}}(\sqrt{E^2-m^2_{\alpha}}R)$ and $\gamma= \frac{\sqrt{E^2-\tilde{k}^2-m^2_{\II}}+i\tilde{k}}{E+m_{\II}}$.
When a localized flux is inserted through the origin, then away from origin, the solutions are still of the form given by (\ref{cap}) with the replacement $\tilde{k}\rightarrow \tilde{k}+\Phi$ for $\Phi$ flux quanta. In this case, square integrability at the origin no longer constrains the coefficients in (\ref{cap}), instead a matching to the solutions in the B-field region must be done. As discussed in the text, the simplest flux profile allowing for analytical solution is to take a delta-function ring~\cite{Alford1989140}: $B(r)=\Phi\delta(r-\epsilon)/(2\pi\epsilon)$. Then inside the ring, normalizability again constraints one coefficient to zero. Matching this at $r=\epsilon$ gives a required relation between the coefficients in the interior of the flux and the rest of the cap. Then the quantization condition is approximately the same as (\ref{cylquant}) with $\tilde{k}\rightarrow \tilde{k}+\Phi$ for $|\tilde{k}+\Phi|> 1/2$. For $|\tilde{k}+\Phi|\leq 1/2$ the explicit form of the B-field must be used to match wavefunctions at $\epsilon$. The result can be summarized by making redefinitions in equation \ref{cylquant} (assuming $\Phi\geq 0$):
$h_{\alpha}= \frac{\sqrt{E^2-m^2_{\alpha}}}{E+m_{\alpha}}(d_{1}J_{\tilde{k}+\frac{1}{2}+\Phi}(\sqrt{E^2-m^2_{\alpha}}R)
-d_{2}J_{-\tilde{k}-\frac{1}{2}-\Phi}(\sqrt{E^2-m^2_{\alpha}}R))$,
$g_{\alpha}= d_{1}J_{\tilde{k}-\frac{1}{2}+\Phi}(\sqrt{E^2-m^2_{\alpha}}R)
+d_{2}J_{-\tilde{k}+\frac{1}{2}-\Phi}(\sqrt{E^2-m^2_{\alpha}}R)$ where 

$d_1=\frac{J_{\tilde{k}-1/2}-d_2 J_{-\tilde{k}+1/2-\Phi}     }{J_{\tilde{k}-1/2+\Phi}}$ $d_2=\frac{J_{\tilde{k}-1/2+\Phi}J_{\tilde{k}+1/2}-J_{\tilde{k}+1/2+\Phi}J_{\tilde{k}-1/2} }{J_{-\tilde{k}-1/2-\Phi}J_{\tilde{k}-1/2+\Phi}-J_{-\tilde{k}+1/2-\Phi}J_{\tilde{k}+1/2+\Phi}}$ with all unspecified $J$s evaluated at $\sqrt{E^2-M_{\alpha}^2}\epsilon$ so long as $\Phi$ is not an integer. For $\Phi$ integer, one further must replace $J_{-\tilde{k}\dots}$ to $Y$ (the second Bessel function) and $-d_2$ to $+d_2$ in $h_{\alpha}$.
%\nocite{*}

%\bibliographystyle{apsrev4-1}
%\bibliography{ACCS}

\section*{References}
\bibliographystyle{iopart-num}
\bibliography{accs}

\providecommand{\newblock}{}
\begin{thebibliography}{10}
\expandafter\ifx\csname url\endcsname\relax
  \def\url#1{{\tt #1}}\fi
\expandafter\ifx\csname urlprefix\endcsname\relax\def\urlprefix{URL }\fi
\providecommand{\eprint}[2][]{\url{#2}}

\bibitem{Jackiw:1984ji}
Jackiw R 1984 {\em Phys. Rev.\/} {\bf D29} 2375

\bibitem{Niemi:1983rq}
Niemi A~J and Semenoff G~W 1983 {\em Phys. Rev. Lett.\/} {\bf 51} 2077

\bibitem{Redlich}
Redlich A~N 1984 {\em Phys. Rev. D\/} {\bf 29}(10) 2366--2374
  \urlprefix\url{http://link.aps.org/doi/10.1103/PhysRevD.29.2366}

\bibitem{PhysRevB.23.5632}
Laughlin R~B 1981 {\em Phys. Rev. B\/} {\bf 23} 5632--5633

\bibitem{PhysRevB.25.2185}
Halperin B~I 1982 {\em Phys. Rev. B\/} {\bf 25} 2185--2190

\bibitem{Nielsen1983389}
Nielsen H and Ninomiya M 1983 {\em Physics Letters B\/} {\bf 130} 389 -- 396
  ISSN 0370-2693
  \urlprefix\url{http://www.sciencedirect.com/science/article/pii/0370269383915290}

\bibitem{PhysRevB.73.125411}
Peres N~M~R, Guinea F and Castro~Neto A~H 2006 {\em Phys. Rev. B\/} {\bf 73}
  125411

\bibitem{PhysRevLett.95.146801}
Gusynin V~P and Sharapov S~G 2005 {\em Phys. Rev. Lett.\/} {\bf 95} 146801

\bibitem{PhysRevLett.98.106803}
Fu L, Kane C~L and Mele E~J 2007 {\em Phys. Rev. Lett.\/} {\bf 98} 106803

\bibitem{PhysRevB.76.045302}
Fu L and Kane C~L 2007 {\em Phys. Rev. B\/} {\bf 76} 045302

\bibitem{Bernevig15122006}
Bernevig B~A, Hughes T~L and Zhang S~C 2006 {\em Science\/} {\bf 314}
  1757--1761

\bibitem{PhysRevLett.96.106802}
Bernevig B~A and Zhang S~C 2006 {\em Phys. Rev. Lett.\/} {\bf 96}(10) 106802
  \urlprefix\url{http://link.aps.org/doi/10.1103/PhysRevLett.96.106802}

\bibitem{Konig02112007}
K\"{o}nig M, Wiedmann S, Br\"{u}ne C, Roth A, Buhmann H, Molenkamp L~W, Qi X~L
  and Zhang S~C 2007 {\em Science\/} {\bf 318} 766--770 (\textit{Preprint}
  \eprint{http://www.sciencemag.org/content/318/5851/766.full.pdf})
  \urlprefix\url{http://www.sciencemag.org/content/318/5851/766.abstract}

\bibitem{hsieh-2008-452}
Hsieh D, Qian D, Wray L, Xia Y, Hor Y~S, Cava R~J and Hasan M~Z 2008 {\em
  NATURE\/} {\bf 452} 970 \urlprefix\url{doi:10.1038/nature06843}

\bibitem{xia-2009-5}
Xia Y, Qian D, Hsieh D, Wray L, Pal A, Lin H, Bansil A, Grauer D, Hor Y~S, Cava
  R~J and Hasan M~Z 2009 {\em NATURE PHYSICS\/} {\bf 5} 398
  \urlprefix\url{doi:10.1038/nphys1274}

\bibitem{Zhang_Liu_Qi_Dai_Fang_Zhang_2009}
Zhang H, Liu C~X, Qi X~L, Dai X, Fang Z and Zhang S~C 2009 {\em Nature
  Physics\/} {\bf 5} 438--442
  \urlprefix\url{http://www.nature.com/doifinder/10.1038/nphys1270}

\bibitem{PhysRevB.78.195424}
Qi X~L, Hughes T~L and Zhang S~C 2008 {\em Phys. Rev. B\/} {\bf 78}(19) 195424
  \urlprefix\url{http://link.aps.org/doi/10.1103/PhysRevB.78.195424}

\bibitem{PhysRevLett.103.196804}
Lee D~H 2009 {\em Phys. Rev. Lett.\/} {\bf 103}(19) 196804
  \urlprefix\url{http://link.aps.org/doi/10.1103/PhysRevLett.103.196804}

\bibitem{Chen06082010}
Chen Y~L, Chu J~H, Analytis J~G, Liu Z~K, Igarashi K, Kuo H~H, Qi X~L, Mo S~K,
  Moore R~G, Lu D~H, Hashimoto M, Sasagawa T, Zhang S~C, Fisher I~R, Hussain Z
  and Shen Z~X 2010 {\em Science\/} {\bf 329} 659--662

\bibitem{Rosenberg}
Rosenberg G, Guo H~M and Franz M 2010 {\em Phys. Rev.\/} {\bf B82} 041104

\bibitem{PhysRevB.84.085312}
Chu R~L, Shi J and Shen S~Q 2011 {\em Phys. Rev. B\/} {\bf 84}(8) 085312
  \urlprefix\url{http://link.aps.org/doi/10.1103/PhysRevB.84.085312}

\bibitem{0953-8984-24-1-015004}
Zhang Y~Y, Wang X~R and Xie X~C 2012 {\em Journal of Physics: Condensed
  Matter\/} {\bf 24} 015004
  \urlprefix\url{http://stacks.iop.org/0953-8984/24/i=1/a=015004}

\bibitem{Gonzalez1993771}
Gonz\'{a}lez J, Guinea F and Vozmediano M 1993 {\em Nuclear Physics B\/} {\bf
  406} 771 -- 794 ISSN 0550-3213
  \urlprefix\url{http://www.sciencedirect.com/science/article/pii/055032139390009E}

\bibitem{PhysRevB.83.075424}
Parente V, Lucignano P, Vitale P, Tagliacozzo A and Guinea F 2011 {\em Phys.
  Rev. B\/} {\bf 83}(7) 075424
  \urlprefix\url{http://link.aps.org/doi/10.1103/PhysRevB.83.075424}

\bibitem{PhysRevB.76.165409}
de~Juan F, Cortijo A and Vozmediano M~A~H 2007 {\em Phys. Rev. B\/} {\bf
  76}(16) 165409
  \urlprefix\url{http://link.aps.org/doi/10.1103/PhysRevB.76.165409}

\bibitem{Moore}
Cho G~Y and Moore J~E 2011 {\em Annals of Physics\/} {\bf 326} 1515 -- 1535
  ISSN 0003-4916
  \urlprefix\url{http://www.sciencedirect.com/science/article/pii/S0003491610002253}

\bibitem{Zhang}
Zhang Y, Ran Y and Vishwanath A 2009 {\em Phys. Rev. B\/} {\bf 79}(24) 245331
  \urlprefix\url{http://link.aps.org/doi/10.1103/PhysRevB.79.245331}

\bibitem{2010arXiv1005.3542Z}
Zhang Y and Vishwanath A 2010 {\em Phys. Rev. Lett.\/} {\bf 105}(20) 206601
  \urlprefix\url{http://link.aps.org/doi/10.1103/PhysRevLett.105.206601}

\bibitem{Alford1989140}
Alford M~G, March-Russell J and Wilczek F 1989 {\em Nuclear Physics B\/} {\bf
  328} 140 -- 158 ISSN 0550-3213
  \urlprefix\url{http://www.sciencedirect.com/science/article/B6TVC-4718HPM-1N/2/c3b4da8f6876777653b6927f42ab7718}

\bibitem{PhysRevD.16.1052}
Jackiw R and Rebbi C 1977 {\em Phys. Rev. D\/} {\bf 16} 1052--1060

\bibitem{PhysRevLett.105.190404}
Bermudez A, Mazza L, Rizzi M, Goldman N, Lewenstein M and Martin-Delgado M~A
  2010 {\em Phys. Rev. Lett.\/} {\bf 105}(19) 190404
  \urlprefix\url{http://link.aps.org/doi/10.1103/PhysRevLett.105.190404}

\bibitem{PhysRevLett.100.096407}
Fu L and Kane C~L 2008 {\em Phys. Rev. Lett.\/} {\bf 100}(9) 096407
  \urlprefix\url{http://link.aps.org/doi/10.1103/PhysRevLett.100.096407}

\bibitem{0201503972}
Peskin M~E and Schroeder D~V 1995 {\em An Introduction To Quantum Field Theory
  (Frontiers in Physics)\/} (Westview Press)

\bibitem{Spinors}
{Penrose} R and {Rindler} W 1988 {\em {Spinors and Space-Time}\/} (Cambridge
  University Press)

\bibitem{PhysRevD.13.3398}
Jackiw R and Rebbi C 1976 {\em Phys. Rev. D\/} {\bf 13} 3398--3409

\bibitem{Nakahara:2003}
Nakahara M 2003 {\em Geometry, Topology and Physics\/} 2nd ed (CRC Press)

\bibitem{Eguchi:1980jx}
Eguchi T, Gilkey P~B and Hanson A~J 1980 {\em Phys. Rept.\/} {\bf 66} 213

\end{thebibliography}
%\begin{thebibliography}{10}
%\bibitem{book1} Goosens M, Rahtz S and Mittelbach F 1997 {\it The \LaTeX\ Graphics Companion\/} 
%(Reading, MA: Addison-Wesley)
%\bibitem{eps} Reckdahl K 1997 {\it Using Imported Graphics in \LaTeX\ } (search CTAN for the file `epslatex.pdf')
%\end{thebibliography}

\end{document}